\documentclass[fleqn,usenatbib]{mnras}

\usepackage{newtxtext,newtxmath}
\usepackage[T1]{fontenc}

\DeclareRobustCommand{\VAN}[3]{#2}
\let\VANthebibliography\thebibliography
\def\thebibliography{\DeclareRobustCommand{\VAN}[3]{##3}\VANthebibliography}

\usepackage{graphicx}	% Including figure files
\usepackage{tabularx}

\newcommand{\xion}{\bar{x}_{\rm i}}
\newcommand{\dt}{\mathrm{d}t}
\newcommand{\nH}{\bar{n}_{\rm H}}
\newcommand{\fcoll}{f_{\rm coll}}
\newcommand{\dfcolldt}{$\mathrm{d}\fcoll/\dt$}

\graphicspath{{./}{figures/}}

\title[Constraining Pop-III models with CMB]{The high-redshift tail of stellar reionization in LCDM is beyond the reach of the low-$\ell$ CMB}

\author[X. H. Wu et al.]{
Xiaohan Wu$^{1}$\thanks{E-mail: xiaohan.wu@cfa.harvard.edu}\href{https://orcid.org/0000-0003-2061-4299}{\includegraphics[scale=0.05]{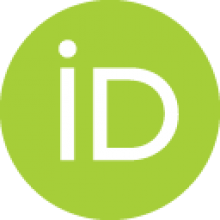}},
Matthew McQuinn$^{2}$,
Daniel Eisenstein$^{1}$,
Vid Ir\v si\v c$^{3,4}$
\\
$^{1}$Harvard-Smithsonian Center for Astrophysics, 60 Garden Street, Cambridge 02138, MA, USA \\
$^{2}$Astronomy Department, University of Washington, Seattle, WA 98195, USA \\
$^{3}$Kavli Institute for Cosmology, University of Cambridge, Madingley Road, Cambridge CB3 0HA, UK \\
$^{4}$Cavendish Laboratory, University of Cambridge, 19 J. J. Thomson Ave., Cambridge CB3 0HE, UK
}

\date{Accepted 2021 September 26. Received 2021 September 3; in original form 2021 May 19}

\pubyear{2021}

\begin{document}
\label{firstpage}
\pagerange{\pageref{firstpage}--\pageref{lastpage}}
\maketitle

\begin{abstract}
The first generation (Pop-III) stars can ionize 1-10\% of the universe by $z=15$, when the metal-enriched (Pop-II) stars may contribute negligibly to the ionization.  This low ionization tail might leave detectable imprints on the large-scale CMB E-mode polarization.  However, we show that physical models for reionization are unlikely to be sufficiently extended to detect any parameter beyond the total optical depth through reionization.  This result is driven in part by the total optical depth inferred by {\it Planck}, indicating a reionization midpoint around $z=8$, which in combination with the requirement that reionization completes by $z\approx 5.5$ limits the amplitude of an extended tail.  To demonstrate this, we perform semi-analytic calculations of reionization including Pop-III star formation in minihalos with Lyman-Werner feedback.  We find that standard Pop-III models need to produce very extended reionization at $z>15$ to be distinguishable at 2-$\sigma$ from Pop-II-only models, assuming a cosmic variance-limited measurement of the low-$\ell$ EE power spectrum.  However, we show that unless there is (1) a late-time quenching mechanism such as from strong X-ray feedback or (2) some other extreme Pop-III scenario, structure formation makes it quite challenging to produce high enough Thomson scattering optical depth from $z>15$, $\tau(z>15)$, and still be consistent with other observational constraints on reionization.
% \tmp{We comment on the \citet{planck18} and \citet{heinrich21} $\tau(z>15)$ constraints which seemingly suggest more information in $\tau(z>15)$ than our finding that the {\it Planck} likelihood is primarily sensitive to the total optical depth through reionization.}
\end{abstract}


%%%%%%%%%%%%%%%%%%%%%%%%%%%%%%%%%%%%%%%%%%%%%%%%%%

%%%%%%%%%%%%%%%%% BODY OF PAPER %%%%%%%%%%%%%%%%%%

\section{Introduction}
\label{sec:intro}

Reionization adds anisotropy to the large-scale CMB E-mode polarization through Thomson scattering the incident primoridal temperature quadrupole, with the anisotropy on scales corresponding to the horizon at the time of this scattering.  The angular scale and width of the ``reionization bump'' in the E-mode power spectrum thus carry coarse-grained information about the global reionization history of the universe.  The parameter this signal constrains most easily is the integrated Thomson scattering optical depth through reionziation, $\tau$.  $\tau$ is mainly sensitive to the redshift of reionization instead of the duration, since coherence on horizon scales indicates that the duration of reionization has to reach thousands of comoving Mpc (e.g. the distance from $z=5$ to 20) to significantly affect the shape of the reionization bump.  However, ionization at $z\gtrsim15$ contributes additional structure to the EE power spectrum at $\ell\sim10-20$  \citep[e.g.][]{heinrich17, heinrich18}.  Thus, the EE power spectrum is potentially useful for constraining star formation at $z\gtrsim15$, times when the main mode of star formation may be Population III \citep[Pop-III][]{kaplinghat03, haiman03, ahn12, ahn20, miranda17, visbal15, qin20cmb}.

%Pop-III stars are only likely to produce a low ionization level in the intergalactic medium (IGM), compared to Pop-II stars in $\gtrsim 10^9\ M_\odot$ halos which are believed to drive a smooth and quick reionization of the universe at lower $z$ owing to exponential growth of $\gtrsim 10^9\ M_\odot$ galaxies with time.  
Starting at $z \gtrsim 20$, Pop-III stars are thought to form in $10^5-10^6\ M_\odot$ halos that have high enough virial temperatures for their H$_2$ to be collisionally excited to cool the gas (\citealt{haiman96, tegmark97, yoshida03}; for recent reviews, see \citealt{bromm13, greif15}).  Since H$_2$ can only cool the gas to $\sim 200$~K, early studies have predicted that these primordial stars should have high masses \citep[$\gtrsim 100\ M_\odot$][]{bromm02, abel02, yoshida06}, although recent works have also found multiple low-mass stars ($\lesssim$ several$\times 10\ M_\odot$) in a halo owing to fragmentation \citep[e.g.][]{susa13, susa14, hirano14, stacy16, sugimura20}.  Massive unenriched stars have high surface temperatures ($\sim 10^5$~K), leading to $\sim10^5$ ionizing photons produced per stellar baryon during a star's lifetime of a few million years \citep{schaerer02}.  This ionizing emission is an order of magnitude more efficient than the metal-enriched Population II (Pop-II) stars ($\sim 4000$ per stellar baryon).  However, the Lyman-Werner (LW) photons generated by the same Pop-III stars quickly form a background owing to their $\sim 100$~comoving Mpc mean free path, which photodissociates H$_2$ and raises the minimum halo mass capable of forming Pop-III stars \citep{haiman97, haiman00, machacek01, yoshida03, wise07, oshea08}.  Pop-III star formation thus self-regulates and the cosmic star formation rate density eventually saturates, reaching a maximum of $10^{-5}-10^{-4}\ M_\odot/$yr/Mpc$^3$ \citep[e.g.][]{visbal18, mebane18}, rates that can ionizes only $1-10\%$ of the $z\gtrsim15$ intergalactic medium (IGM) \citep[e.g.][]{ahn12, ahn20, visbal20}.

Despite this low ionization level, \citet{irsic20} suggested that the $\tau(z>15)<0.006$ 2-$\sigma$ upper limit constraint obtained by \citet{planck18}, where $\tau(z>15)$ is the optical depth contributed by the $z>15$ tail of reionization, is able to rule out a range of parameter space in their Pop-III star formation models.  The study of \citet{irsic20} builds on previous papers that have constrained Pop-III models with CMB data \citep{ahn12, ahn20, miranda17, visbal15, qin20cmb}.  Although in the final stages of preparation of this paper we learned that this $\tau(z>15)<0.006$ 2-$\sigma$ upper limit in the original {\it Planck} analysis has been found to be incorrect and corrected to $\tau(z>15)<0.018$ \citep{planck21correction}, future CMB surveys aim to measure the EE power spectrum at $\ell\sim10-20$ and, hence, $\tau(z>15)$ to much higher precision than {\it Planck} \citep{watts20}.  LiteBIRD's goal of full-sky observations of primordial B-modes with 2~$\mu$K arcmin white noise level will push the constraining power on the reionization history to the cosmic variance limit \citep{litebird}. 

Motivated by the finding of \citet{irsic20} that a low $\tau(z<15)$ upper limit can constrain models, plus the prospects for improvement with upcoming CMB observatories, our aim is more systematically understand the power of upcoming CMB observations for constraining Pop-III models, using the simplest possible physically motivated models.  To do so, we explore a range of parameter space of highly-uncertain Pop-III modeling.
\emph{While modeling of the early star formation history is inherently messy, our main finding that the CMB is unlikely to constrain Pop-III star formation seems robust to the astrophysical uncertainties.}
We find that assuming our fiducial model of Pop-III star formation and a cosmic variance-limited observation, $\tau(z>15)>0.009$ is needed for a Pop-III model to be distinguished at $>2$-$\sigma$ level from Pop-II-only models using the low-$\ell$ EE power spectrum.  However, when the models are required to match the total optical depth through reionization (total $\tau$) obtained by {\it Planck} and complete reionization by $z=5.5$, none of the models produce such high $\tau(z>15)$; we find that physical models cannot produce very extended reionization while still satisfying the total $\tau$ and endpoint of reionization constraints.  While we phrase our result in terms of the $\tau(z>15)$ required, an implication is that physical models for reionization are unlikely to be sufficiently extended to allow a significant detection of any parameter beyond the total $\tau$.   %most of the Pop-III parameter space is already ruled out by the total $\tau$ and Lyman-$\alpha$ forest constraints on reionization.
We explore possible forms of Pop-III models that are able to produce more plateaued reionization histories or ``double reionization'', which may leave more distinguishable features in the low-$\ell$ CMB.  While we are only able to devise models in which the tail of reionization can ever be detected at $2\sigma$, most of the models needed to generate a detectable signal require feedback or Pop-III star formation to evolve strongly with redshift in a manner that is well outside such evolution in mainstream models.  %We highlight one $2\sigma$-detectable model as plausible, one with relatively less efficient star formation per minihalo compared to our typical models but with very weak Lyman-Werner feedback so that many minihalos host Pop-III stars.  %strengthens our conclusions that the low-$\ell$ CMB is unlikely to be a useful tool for constraining Pop-III models.

% Curiously, the $\tau(z>15)=0.009$ threshold we find that is needed for a future cosmic variance-limited detection of this parameter is close to the 2-$\sigma$ upper limit found by \citet{planck18} of $\tau(z>15)<0.006$, even though the {\it Planck} likelihood is far from being the cosmic variance limit.  The wild fluctuations in the FlexKnot reionization histories \citep{millea18} used in the \citet{planck18} analysis produce larger differences in the CMB EE power spectrum than our physically motivated smooth and monotonic reionization histories.  Additionally, the set of reionization histories consistent with the {\it Planck} total $\tau$ tend to have low $\tau(z>15)$, which we show further tilts the posterior to low values.  Both effects we suggest lead to the surprisingly tight low $\tau(z>15)$ constraint, despite the less constraining {\it Planck} likelihood.  

This paper is organized as follows.  In Section~\ref{sec:methods} we describe our Pop-III star formation models.  We present our calculations of the reionization history and predictions on the CMB EE power spectrum and $\tau(z>15)$ in Section~\ref{sec:main_results}.  We discuss more exotic forms of Pop-III models that are likely to produce more detectable imprints on the CMB in Section~\ref{sec:exotic_models}, %explore possible causes of the low {\it Planck} 2-$\sigma$ upper limit on $\tau(z>15)$ in Section~\ref{sec:planck}, 
and conclude in Section~\ref{sec:conclusions}.

\section{Modeling the ionization history of the universe}
\label{sec:methods}

The mass-averaged ionization fraction $\xion(z)$ determines the amount of Thomson scattering at every redshift. We calculate the time evolution of $\xion$ via\footnote{We note that equation~(\ref{eq:dxidt}) is a simplified treatment of the reionization history and relies on the assumption that sources produce fully ionized bubbles.  However, accounting for the burstiness of the sources may result in larger ionized bubbles and long-lived H{\sc ~ii} regions \citep{hartley16}.  This effect may lead to a different shape of the reionization history that is not captured by equation~(\ref{eq:dxidt}).  However, the clustering of Pop-III sources within ionized regions makes the fully ionized assumption of equation~(\ref{eq:dxidt}) more applicable, as we suspect must occur for the large Pop-III-plateau ionization fractions of $\gtrsim 10\%$ that we need to generate detectable $\tau(>15)$.}
\begin{equation}
\frac{\mathrm{d}\xion}{\mathrm{d}t} = \epsilon - \left( 1+\frac{Y}{4X} \right)\alpha(T_0) C \nH \xion,
\label{eq:dxidt}
\end{equation}
where $\epsilon$ is the ionizing photon emissivity per hydrogen atom determined by the star formation rates (SFR) of Pop-III and Pop-II stars, and $\nH$ is the volume-averaged hydrogen number density.  $X=0.76$ and $Y=1-X$ are the mass fractions of hydrogen and helium respectively.  $C$ is the clumping factor \citep[e.g.][]{pawlik09, finlator12, chen20}, which characterizes the excess recombination in an inhomogeneous IGM compared to a uniform one with temperature $T_0=10^4$~K with a CASE B recombination rate of $\alpha(T_0)=2.6\times10^{-13}$~cm$^{-3}$s$^{-1}$.  Modeling the ionization history of the universe thus requires both a model for star formation and a model for the IGM clumpiness.  We will briefly describe our fiducial model for computing $\xion$ below, and illustrate in more detail the modeling procedure and the parameters involved in Appendix~\ref{sec:appendix_model}.  Given the huge uncertainties in modeling Pop-III star formation, we will focus the discussions in the main text on bracketing the major uncertainties in the modeling, and explore a few model variations in the appendix. %~\ref{sec:appendix_LW} and \ref{sec:appendix_PopII}.
Our model relies on the following ingredients.

\subsection{Pop-III star formation}
\label{sec:methods_popiii}

We have developed extremely flexible models for Pop-III star formation to explore what scenarios could significantly affect the EE power spectrum.
Our fiducial model assumes that Pop-III SFR is proportional to the dark matter accretion rate (also denoted as the ``\dfcolldt'' model), with the coefficient given by the star formation efficiency (SFE).
Since some recent simulations support the picture of several Pop-III stars forming in a halo with a total Pop-III stellar mass of a few hundred $M_\odot$ that does not scale with the halo mass (e.g. \citealt{susa14, stacy16}, but see \citealt{hirano14, hirano15}), we also examine models where a fixed stellar mass forms per halo (denoted as the fixed-stellar-mass model).  However, we do not consider the fixed-stellar-mass model as the fiducial case, mainly because such models when coupled with our fiducial LW feedback scheme are likely not distinguishable from Pop-II-only models using the low-$\ell$ EE power spectrum, as we will show below in Section~\ref{sec:main_results} and \ref{sec:exotic_models}.  Moreover, as simulations supporting the fixed-stellar-mass model have only explored a small halo mass range ($10^5$-$10^6\ M_\odot$), higher mass halos may form more Pop-III stars \citep[e.g.][]{skinner20}.
% Since early works support the picture of one massive Pop-III star forming in a minihalo owing to raidation feedback and lack of fragmentation \citep[e.g.][]{abel02, oshea07}, we also examine models where a fixed stellar mass forms per halo  \citep[e.g. only one Pop-III star per halo][denoted as the fixed-stellar-mass-model]{abel02, wise05, oshea07}.
In the fiducial ``\dfcolldt'' model, a typical value of the SFE can be calculated by assuming $100\ M_\odot$ Pop-III stellar mass forming in a $10^6\ M_\odot$ halo, giving a SFE of $\sim10^{-3}$ \citep{loeb13}, consistent with numerical simulations \citep{skinner20}.  The dark matter accretion rate is given by the rate of change in the cosmic mass fraction collapsed in to halos that form Pop-III stars \citep[e.g.][]{visbal15, irsic20}.  We assume that Pop-III stars form in halos within a mass range $M_{\rm min} < M < M_{\rm a}$, and calculate the collapsed fraction using the Sheth-Tormen mass function \citep{sheth99}.
To explore the Pop-III modeling parameter space, we vary the Pop-III SFE when using the fiducial ``\dfcolldt'' model, or the Pop-III stellar mass formed in each halo when using the fixed-stellar mass model.  In the fixed stellar mass model, once a halo can cool, it forms stars and the subsequent supernovae sterilize the halo from forming subsequent stars.

The minimum halo mass $M_{\rm min}$ of Pop-III star formation is set by the LW background which photo-dissociates molecular hydrogen.  Instead of calculating the LW intensity $J_{\rm LW}$ by fully modeling the ``sawtooth'' shape of the LW attenuation as a function of frequency \citep{haiman00, wise05}, we compute $J_{\rm LW}$ by making the simple ``screening'' assumption that the IGM is transparent to LW photons until they are redshifted into a Lyman series line and absorbed \citep[e.g.][]{visbal15, mebane18}.  We assume that a LW photon can be redshifted by 2 per cent before reaching a Lyman series line, which was found to produce Pop-III SFR consistent with more sophiticated treatments of the LW background \citep{visbal15}.  Given $J_{\rm LW}$, we calculate $M_{\rm min}$ using the fitting formula provided by \citet{machacek01}, which has been used in most semi-analytic models of Pop-III star formation \citep[e.g.][]{fialkov13, visbal15, visbal18, visbal20, mebane18, mebane20, irsic20, qin20i, qin20ii}.
We assume that star formation transforms to the Pop-II channel at a halo mass $M_{\rm a}$ corresponding to a virial temperature $T_{\rm vir}=10^4$~K, above which atomic cooling is efficient \citep[e.g.][]{trenti09, mebane18, irsic20}.  $10^4$~K is also the virial temperature above which LW feedback becomes inefficient \citep{oshea08}.

Since some works have argued for or adopted a stronger LW feedback than the \citet{machacek01} model \citep[e.g.][]{yoshida03, oshea08, ahn12, ahn20, miranda17}, while recent works favor weaker LW feedback owing to H$_2$ self-shielding at high column densities \citep{kulkarni20, schauer20, skinner20}, we manipulate the strength of the LW feedback by changing the amplitude of the fiducial $M_{\rm min}$, or assuming no LW feedback.  We also explore different functional forms of $M_{\rm min}$ following \citet{kulkarni20, schauer20, trenti09}, with a detailed comparison of these models in Appendix~\ref{sec:appendix_LW}.

We do not include the chemical feedback from Pop-III supernova explosion in our modeling.  On the one hand, there has been a lot of debate in the literature about the role of metal enrichment on the quenching of Pop-III star formation \citep{yoshida04, wise12, greif10, muratov13a}, the importance of external and internal enrichment \citep{trenti09b, maio11, hicks20}, and the delay time of Pop-II star formation after supernova explosion \citep{muratov13b, jeon14, smith15}.  These issues are further complicated by the Pop-III stellar mass and the initial mass function which determines the metallicity and energy yield \citep{chiaki18, jaacks18}.
On the other hand, different works seem to agree that Pop-II star formation takes over Pop-III at $z=12-15$ \citep{maio10, jeon15, abe21metal}, which is roughly captured in our calculations (see Section~\ref{sec:main_results}).
% Moreover, \citet{mebane18} found that the assumption of Pop-III to Pop-II transition occurring at $T_{\rm vir}=10^4$~K
We also do not attempt to incorporate the quenching of Pop-III star formation owing to the mechanical feedback of supernova explosions inside adjacent halos \citep{wise08, johnson13, abe21metal}, which is difficult to model in semi-analytic calculations.  Our fixed-stellar-mass models do envision a single star formation epoch in each halo, with feedback disrupting subsequent Pop-III star formation.

\subsection{Pop-II star formation}
\label{sec:methods_popii}

The faster Pop-II stars complete reionization and minimize their contribution to the total $\tau$, the more room there is for a distinct high-$z$ E-mode signal from an early extended Pop-III phase.  Therefore, to explore models with the most distinguishable E-mode signals, we adopt a fiducial model where reionization is quite brief, matching the observed star formation rate density of bright galaxies (see Appendix~\ref{sec:appendix_model}).  We note that such a model appears inconsistent with observations of the Lyman-$\alpha$ forest at $z<6$ that favor a flat or increasing ionizing emissivity with increasing redshift \citep{hm12, becker21} and the picture based on inferences of the emissivity at the end of reionization that reionization was `photon starved' and thus extended \citep{bolton07}.  Specifically, we model Pop-II star formation following \citet{tacchella18}, who assumes that the Pop-II SFR is proportional to the halo accretion rate via a redshift-independent Pop-II SFE proportional to halo mass.  We assume that Pop-II star formation occurs in halos with masses above $M_{\rm a}$ and is not suppressed by photoheating feedback.\footnote{Some studies have suggested that a non-negligible fraction of Pop-II stars may form in halos of $10^{6.5}-10^8\ M_\odot$ owing to metal enrichment by Pop-III stars and can contribute to $\sim30\%$ of the total ionizing photons for completing reionization \citep{wise14, xu16fesc, norman18}.  A lowered atomic cooling halo mass threshold may lead to more extended reionization by Pop-II stars.  However, owing to the uncertainties in modeling metal enrichment as we pointed out above, we simply approximate the transition to Pop-II star formation with an atomic cooling threshold of $10^4$~K, which corresponds to $5\times10^7\ M_\odot$ at $z=15$.  The effect of a more extended Pop-II reionization is also explored by using the \citet{furlanetto17} model in our work.}

While the \citet{tacchella18} model bounds the maximum contribution of Pop-III stars to reionization at $z>15$, we also explore the effects of a much more extended Pop-II model from \citet{furlanetto17} where Pop-II star formation is non-negligible at $z>15$ (see Figure~\ref{fig:popIIsfrd}).  The model assumes that star formation is regulated by momentum conservation of the supernova blastwaves, and leads to a star formation rate density of $3\times10^{-5}\ M_\odot/{\rm yr/Mpc^3}$ at $z=25$.  We will show that such more extended models makes it even harder for Pop-III stars to contribute significantly enough to reionization at $z>15$ to leave detectable imprints on the CMB.

\subsection{Clumping factor}
\label{sec:methods_clumping}

The high-$z$ component of reionization and, hence, the E-mode signal is enhanced the larger the \emph{decrease} in clumping factor with increasing redshift, as this decrease would lead to fewer recombinations and hence larger $x_i$.  Here we argue that the clumping factor $C$ is unlikely to decrease significantly with redshift, and adopt a fixed value of $C=3$.  %At $z\gtrsim15$ when a crucial source of density inhomogeneity comes from the photo-evaporation of minihalos \citep{shapiro04, iliev05}, 
The value $C=3$ is the concordance of simulations focusing on $z<15$ when much of the volume is ionized (\citealt{pawlik09, mcquinn11, shull12, park16} and see especially \citealt{daloisio20}).  To latch onto this concordance low redshift value, there is not a lot of room for $C$ to decrease to high redshift, as it seems implausible that it could evolve to significantly less than unity (indicating the ionization of void regions).  The value $C=3$ is further consistent with the findings of numerical works on Pop-III reionization \citep[e.g. a few to 10 in][]{alvarez06, wise08, wise14}, although some works do include a factor of two decrease in $C$ from $z\sim5$ to  $z\sim30$ that owes to structure formation \citep[e.g.][]{visbal18, ahn20}. We show in Section~\ref{sec:exotic_models} that a decrease that scales as $C\propto (1+z)^{-8}$ is required to significantly change our results. 

However, we think it is most likely that the clumping may scale in the opposite direction, of increasing with higher redshift, reducing the high-$z$ polarization signal.  The small global ionized fraction at $z>15$ makes it more common for the ionizing regions not to percolate such that Pop-III stars mostly ionize their immediate surroundings.  These surroundings are overdense, their $\lesssim5$~proper kpc HII regions corresponding to overdensities of $\delta = 6-10$  \citep{whalen04, kitayama04, abel07}, and, even ignoring inhomogeneities within an ionized bubble the clumping factor would be $C = 1+\delta$.  This effect has not be captured in previous works and we defer to future work with higher resolution simulations to thoroughly study the IGM clumping around Pop-III star-forming halos.

We also note that the ionization at high redshifts is shaped by SFR$/C$ (eqn.~\ref{eq:sfr_estimate}) and so our flexible prescription for SFR compensates for our rigid prescription for $C$.\\

At each time step with a given ionized fraction, we calculate $J_{\rm LW}$ and the Pop-III SFR by looping over these quantities and $M_{\rm min}(J_{\rm LW})$ iteratively until convergence, assuming that Pop-III and Pop-II stars produce $1\times 10^5$ and $9700$ LW photons per stellar baryon respectively \citep[e.g.][]{schaerer02, mebane18, irsic20}.  We then calculate $\xion$ at the next time step using the resulting star formation rates, taking the number of ionizing photons produced per stellar baryon to be $6\times 10^4$ and $4\times10^3$ in Pop-III and Pop-II stars respectively \citep[e.g.][]{schaerer02, alvarez06, visbal15}.  We assume that all ionizing photons escape from a Pop-III star forming halo \citep[e.g.][]{whalen04, abel07}, and vary the constant ionizing photon escape fraction in Pop-II halos as a free parameter.  While it would be unexpected for the escape fraction to evolve substantially at a given halo mass over the brief reionization phase, some models do invoke an escape fraction that decreases with halo mass \citep[e.g.][]{finkelstein19}.  Such a decrease would lengthen the duration of the Pop-II phase of  reionization and strengthen our results.

Finally, given the ionization history of the universe, we compute the Thomson scattering optical depth $\tau$ by integrating the electron number density along the line of sight, assuming that helium is singly ionized before HeII reionization at $z=3$ and fully ionized after that.  We obtain the CMB EE power spectrum for each reionization history using CLASS \citep{classii}.  In what follows, we will quote the $\Delta\chi^2$ values of the EE power spectra, computed using the EE power at $\ell=2-100$ assuming cosmic variance limited.  We calculate this $\Delta\chi^2$ value for each model with Pop-III star formation against a model involving only the Pop-II contribution to reionization with the same total $\tau$.\footnote{We note that we have fixed the amplitude of scalar fluctuations $A_s$ when generating the EE power spectrum, instead of fixing $\exp(-2\tau)A_s$.  Implementing the latter only results in minor changes in the $\Delta\chi^2$ values since we compare models with the same total $\tau$.}

\section{The constraining power of large-scale CMB E-mode polarization on Pop-III models}
\label{sec:main_results}

\begin{figure*}
\includegraphics[width=2\columnwidth]{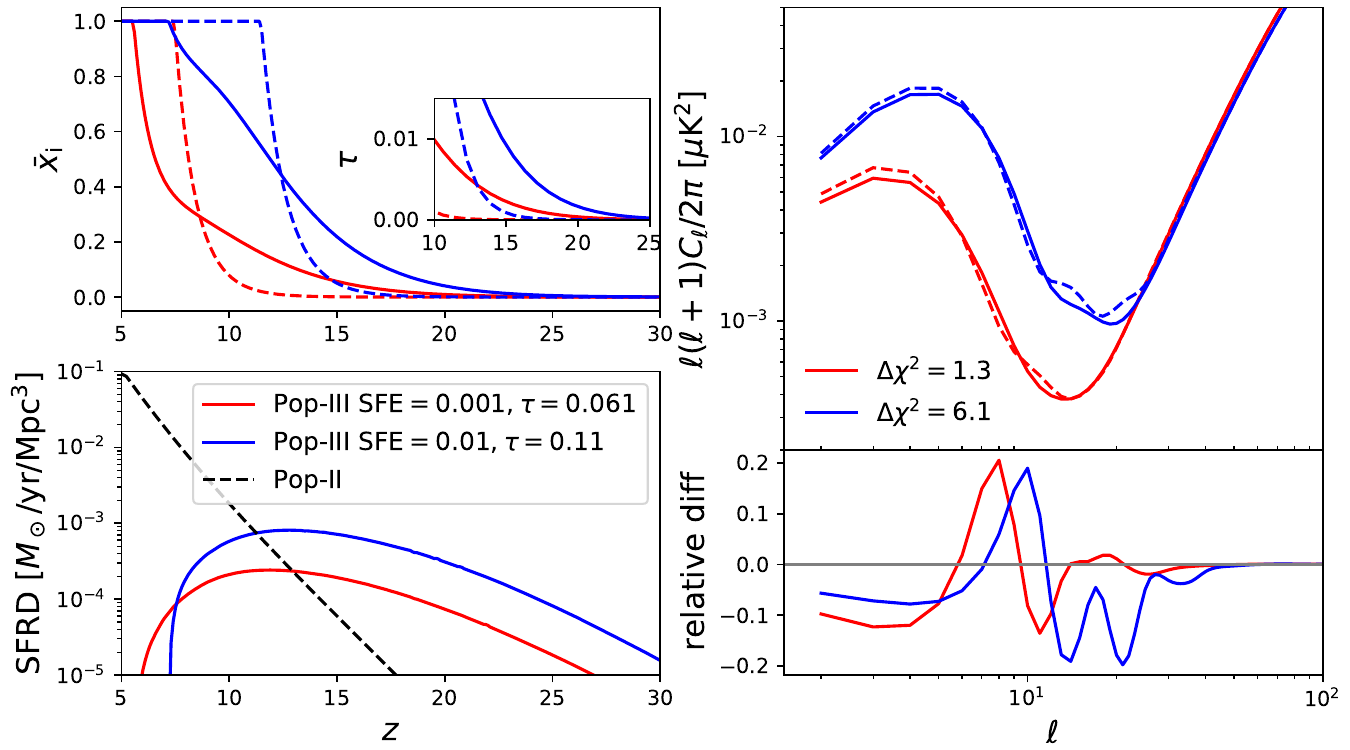}
\caption{A comparison of the fiducial Pop-III models (solid lines) against the Pop-II-only models (dashed lines) with the same total $\tau$ as their Pop-III counterparts.  Red and blue represent models using Pop-III star formation efficiencies of $0.001$ and $0.01$ respectively.  The former gives a reionization history consistent with the {\it Planck} total $\tau$ constraint, and the latter produces a total $\tau$ that is too high.  The top-left panel shows the reionization histories, and the inset illustrate the accumulated $\tau$ at $z=10-25$.  The bottom-left panel presents the SFR densities of Pop-III (solid lines) and Pop-II (black dashed line).  The right panels show the EE power spectra (top panel) and the relative differences of the Pop-III models to their Pop-II-only counterparts (bottom panel).  The Pop-III model consistent with the {\it Planck} total $\tau$ constraint only results in $\tau(z>15)=0.002$ and $\Delta\chi^2=1.3$ for a cosmic variance-limited EE power spectrum measurement over $\ell=2-100$, while the history producing $\tau(z>15)=0.009$ and $\tau=0.011$ gives $\Delta\chi^2=6$.}
\label{fig:fidLW_Cl}
\end{figure*}

We now turn to the exploration of the model and its implications for the large-scale CMB E-mode polarization.  We first illustrate our modeling and analysis procedure.  Figure~\ref{fig:fidLW_Cl} compares the following quantities for two fiducial Pop-III models and their corresponding Pop-II-only models with the same total $\tau$: the reionization histories and the accumulated $\tau$ (top-left and its inset), star formation rate densities (bottom-left), the EE power spectra (top-right) created with CLASS \citep{classii}, and the relative differences between the EE power spectra of the Pop-III and Pop-II-only model (bottom-right).  The red and blue solid lines represent Pop-III models using star formation efficiencies of $0.001$ and $0.01$, producing total $\tau=0.061, 0.11$ and $\tau(z>15)=0.002, 0.009$ respectively.  The red and blue dashed lines show the corresponding Pop-II-only models with the same total $\tau$, and the black dashed line in the bottom-left panel illustrates the Pop-II SFR density.  The top-right panel also indicates the cosmic variance-limited $\Delta\chi^2$ between each Pop-III model and its corresponding Pop-II model.

The Pop-III model with a SFE of $0.001$ produces negligible differences in the EE power at $\ell=10-30$ from its Pop-II-only counterpart owing to the low $\tau(z>15)=0.002$, leading to $\Delta\chi^2=1.3$.  On the other hand, the Pop-III model with a SFE of $0.01$ which is near the maximum efficiency that studies have found for Pop-III models \citep[e.g.][]{trenti09, visbal15} results in $\tau(z>15)=0.009$ and $10-20\%$ differences in the EE power at $\ell=10-30$ from the Pop-II-only model, giving $\Delta\chi^2=6$.\footnote{The much shallower and extended Pop-III star formation histories produce more prolonged ionization at $z>15$ than Pop-II, but the resulting EE power at $\ell=10-30$ is not necessarily higher than what the Pop-II-only models generate, owing to the more rapid reionization histories of the Pop-II-only models creating ringing into the power spectra \citep{hu03}.}  However, regardless of whether such high star formation rates are possible, the model produces too high total $\tau$ compared to the \citet{planck18} observations.  This model is therefore ruled out by the low total $\tau$ constraint, even though it can be distinguished at $2.4$-$\sigma$ from a Pop-II-only model with the same total $\tau$ using the low-$\ell$ CMB EE power spectrum.  Our results suggest that Pop-III models consistent with the {\it Planck} low total $\tau$ are unlikely to give high enough $\tau(z>15)$ to be detectable through the low-$\ell$ CMB.\\

\begin{figure*}
\includegraphics[width=2\columnwidth]{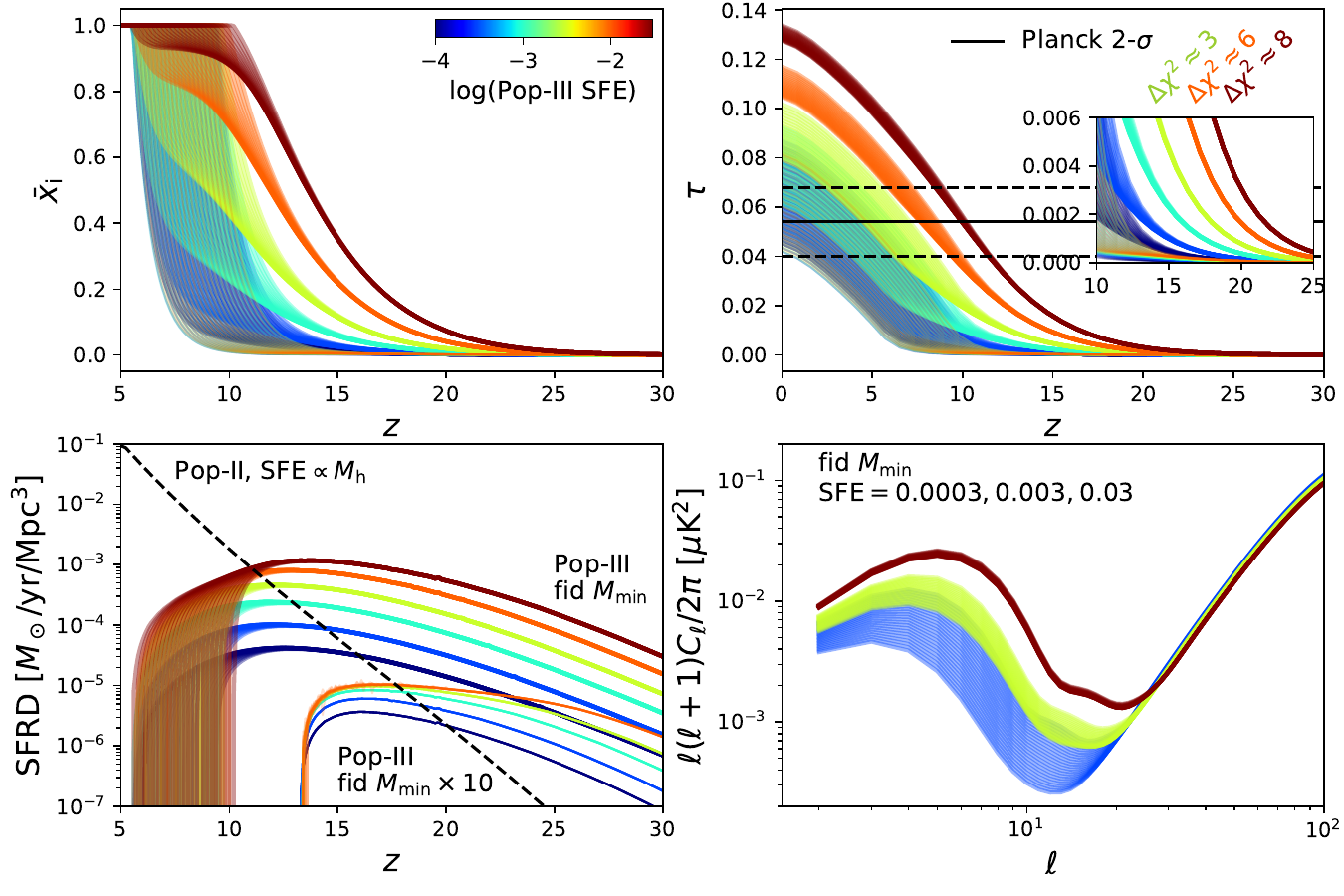}
\caption{Results of our calculations using the fiducial model and the model variation with the $10 \times M_{\rm min}$ strong LW feedback.  We vary the Pop-III star formation efficiency with 6 values from $10^{-4}$ to $0.03$ and the Pop-II escape fraction, and only show models that completes reionization by $z=5.5$.  Owing to instabilities in the calculations when LW feedback is strong, we do not show results with $10\times M_{\rm min}$ and Pop-III SFE of $0.03$.  The curves are color-coded by log of the Pop-III star formation efficiency.  For each series of curves with a single Pop-III SFE, the spread in the curves owes to variations in the Pop-II escape fraction.  The thicker and thinner sets of curves represent models with the fiducial $M_{\rm min}$ and using $10 \times M_{\rm min}$ respectively, where the thin curves are mostly overlapped by the thick ones except in the bottom left panel.  The top-left panel shows the ionization histories.  The bottom-left panel presents the Pop-III (solid lines) and Pop-II (black dashed line) SFR densities as a function of $z$.  The top-right panel illustrates the accumulated optical depth $\tau$ and the black lines show the {\it Planck} 2-$\sigma$ constraints on the total $\tau$.  The inset shows a zoom-in of $\tau$ in the redshift range 10-25.  The $\Delta\chi^2$ values for a cosmic variance-limited EE power spectrum measurement, obtained by comparing the EE power at $\ell=2-100$ of each Pop-III model to a Pop-II-only model with the same total $\tau$, are indicated for the 3 sets of curves with the fiducial $M_{\rm min}$ and the largest Pop-III SFE.  The bottom-right panel illustrates the EE power spectra for models using the fiducial $M_{\rm min}$ and ${\rm SFE} = 0.0003, 0.003, 0.03$.}
\label{fig:model_Sandro_master}
\end{figure*}

We now examine predictions from our fiducial Pop-III model with a comprehensive exploration of the modeling parameter space by exploring different Pop-III SFE and the amplitude of $M_{\rm min}$.  Figure~\ref{fig:model_Sandro_master} shows the results of our calculations using the fiducial $M_{\rm min}$ and $10 \times M_{\rm min}$, represented by the thicker and thinner sets of curves respectively.  Results using $10\times M_{\rm min}$ are almost all overlapped by those using the fiducial $M_{\rm min}$ except in the bottom left panel.  The curves are color-coded by log of the Pop-III star formation efficiency, which we vary with 6 values: $10^{-4}, 3\times10^{-4}, 10^{-3}, 3\times10^{-3}, 10^{-2}, 3\times10^{-2}$.  For each combination of SFE and LW feedback strength, we calculate a set of reionization histories with a range of Pop-II ionizing photon escape fractions listed in Table~\ref{tab:params}, with an upper limit of 10 to allow for more contribution from Pop-II stars to ionization at $z>15$.  The Pop-II escape fraction only has a small effect on the Pop-III SFR through small changes in the high-$z$ ionzied fraction.  We only keep the models that complete reionization before $z=5.5$, consistent with observations of the Lyman-$\alpha$ forest \citep[e.g.][]{fan06, mcgreer15, becker15}.  Owing to instabilities in the calculations when LW feedback is strong, we do not show results with $10\times M_{\rm min}$ and Pop-III SFE of $0.03$.  Top-left panel shows the ionization histories.  Bottom-left panel presents the Pop-III (solid lines) and Pop-II (black dashed line) SFR densities as a function of $z$.  Top-right panel illustrates the accumulated optical depth $\tau$ and the black lines show the {\it Planck} 2-$\sigma$ constraints on the total $\tau$.  The inset shows a zoom-in of $\tau$ in the redshift range 10-25.  The $\Delta\chi^2$ values of the EE power spectra, obtained by comparing the EE power at $\ell=2-100$ of each Pop-III model to a Pop-II-only model with the same total $\tau$, are indicated for the 3 sets of curves with the fiducial $M_{\rm min}$ and the largest Pop-III SFE.  Bottom-right panel illustrates the EE power spectra for models using the fiducial $M_{\rm min}$ and ${\rm SFE} = 0.0003, 0.003, 0.03$.

Using $10 \times M_{\rm min}$ leads to suppressed Pop-III star formation rates and negligible $\tau(z>15)$.  This implies that Pop-III models which are distinguishable from Pop-II-only ones at $>2$-$\sigma$ level using the low-$\ell$ EE power spectrum cannot have strong LW feedback, even though such models may show plateaued Pop-III SFR densities at high Pop-III SFE owing to the self-regulation of Pop-III star formation.  This self-regulation only occurs in the case of strong LW feedback when the increased LW background owing to the rising Pop-III SFE brings the minimum halo mass for star formation close to our imposed maximum halo mass ($T_{\rm vir}=10^4$~K).  %We will therefore focus on results using the fiducial $M_{\rm min}$ below.
% \vid{What about weaker LW feedback? Since this is answered in the next section you could refer to that (Kulkarni+20) model.}

Using the fiducial $M_{\rm min}$, models with Pop-III SFE $\ge0.01$ leads to $\tau(z>15)>0.006$, which the {\it Planck} 2-$\sigma$ upper limit on $\tau(z>15)$ is able to rule out.  However, when required to complete reionization by $z=5.5$ and have total $\tau$ consistent with the {\it Planck} constraint within 2-$\sigma$, models with Pop-III SFE $\gtrsim0.003$ are all ruled out owing to high total $\tau$.  This conclusion remains unchanged if we allow reionization to complete by $z=5$ instead, which decreases the total $\tau$ lower bounds by a small fraction.  This implies that the total $\tau$ and endpoint of reionization constraint imposed by the Lyman-$\alpha$ forest observations are already able to restrict the allowable Pop-III parameter space \citep[see also][]{visbal15}, before further taking into account $\tau(z>15)$ and the CMB EE power spectrum at $\ell\sim10-20$.  The {\it Planck} $\tau(z>15)$ constraint thus does not help putting more constraints on Pop-III models.%\vid{How important is the $z>5.5$ constraint? What happens if you use $z=5.0$?}
%This leads to the question of whether future CMB surveys, which can likely push down the $\tau(z>15)$ upper limit, can be useful in constraining Pop-III models.  However, our results show that models with Pop-III ${\rm SFE} \le 0.001$ which gives $\tau(z>15)<0.003$ only result in $\Delta\chi^2 < 3$.  Only models with ${\rm SFE} \gtrsim 0.01$ produces $\Delta\chi^2 > 4$ and thus can be distinguishable from Pop-II only models at $>2$-$\sigma$ level with future CMB data, and such models lead to high $\tau(z>15)$ as well.  The contribution of the large-scale EE power spectrum to constraining Pop-III models is thus likely very limited.

\begin{figure*}
\begin{center}
\includegraphics[width=0.52\linewidth]{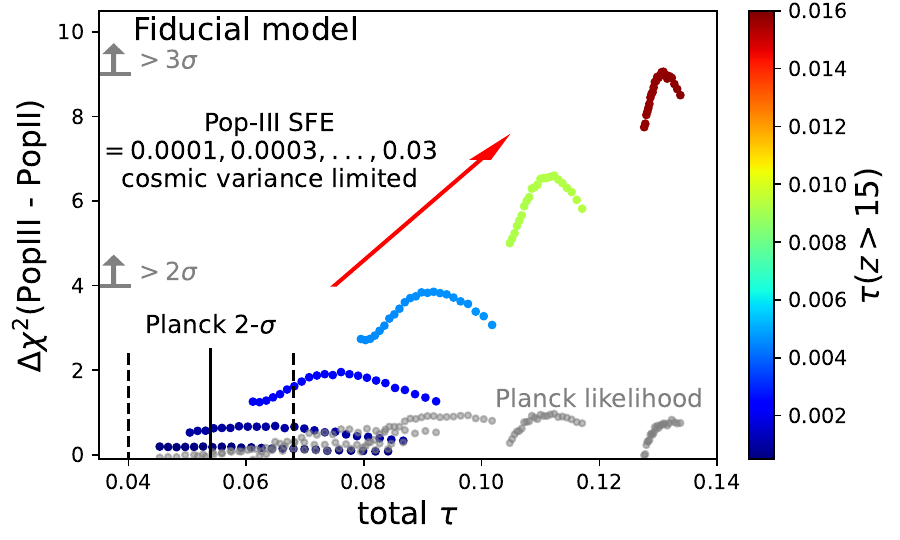}
% \end{center}
% \begin{center}
\includegraphics[width=0.42\linewidth]{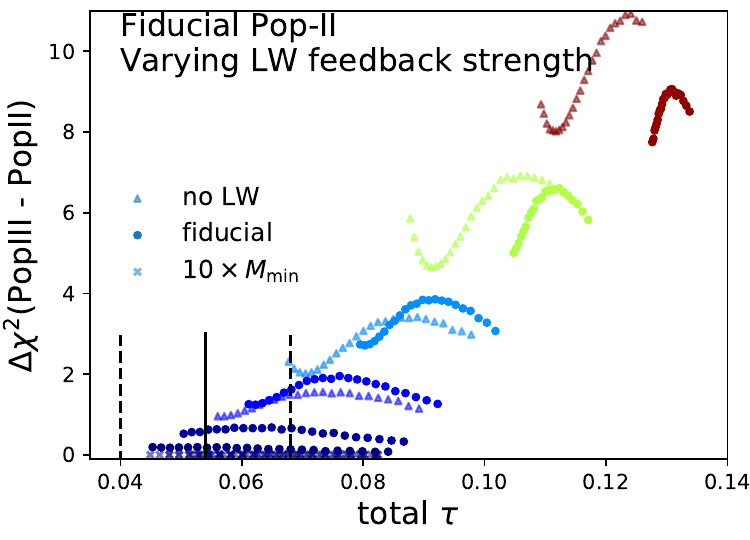}
\includegraphics[width=0.42\linewidth]{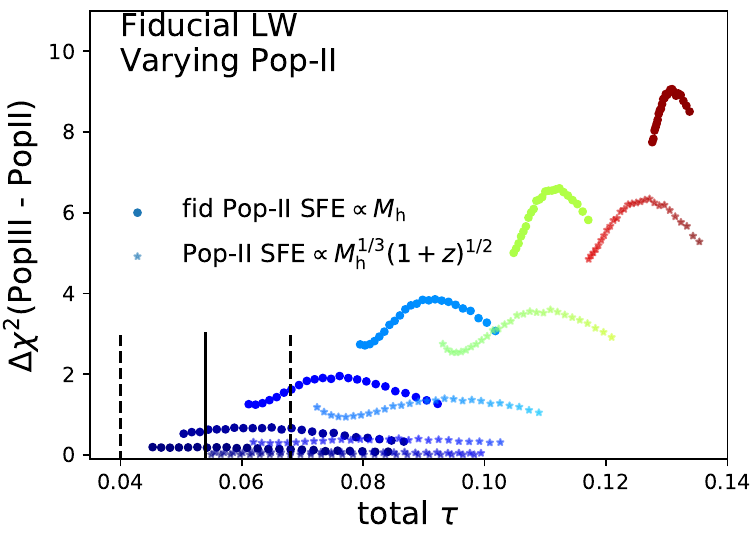}
\hspace{1.5 cm}
\includegraphics[width=0.42\linewidth]{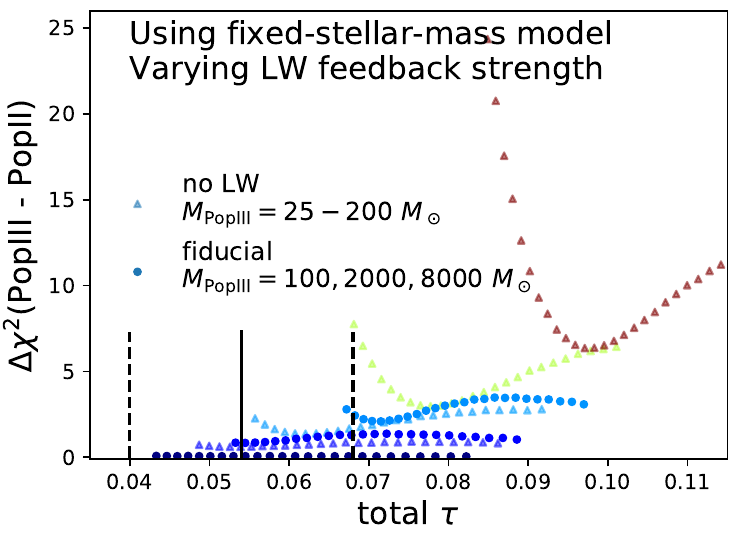}
\end{center}
\caption{The panels show the $\Delta\chi^2$ values of the EE power spectrum between Pop-III models and Pop-II-only models with the same total $\tau$ for the fiducial model and 3 sets of model variations.  The points are color-coded by the values of $\tau(z>15)$, which are determined by the Pop-III SFE or stellar mass.  The $\Delta\chi^2$ values are a strong function of $\tau(z>15)$, a primary result of this study.  The top-left panel illustrates results for the fiducial model.  The vertical black lines show the {\it Planck} 2-$\sigma$ constraint on the total $\tau$.  For comparison (and only in the top-left panel), we also show the effective $\Delta\chi^2$ values using the {\it Planck} {\tt SimAll} likelihood in gray, defined as twice the difference between the log likelihoods of each Pop-III and Pop-II-only pair with the same total $\tau$.  The top-right panel presents our calculations of changing the LW feedback strength, where dots, triangles, and crosses (all piled at the bottom) represent the fiducial $M_{\rm min}$, no LW, and $10\times M_{\rm min}$ respectively.  The bottom-left panel explores the effects using the \citet{furlanetto17} Pop-II model (stars) against the fiducial \citet{tacchella18} model (dots).  The bottom-right panel shows results of using the fixed-stellar-mass model with different Pop-III stellar masses assuming the fiducial LW feedback (dots) and no LW (triangles), where the fiducial LW feedback barely leads to $\Delta\chi^2>4$ even using a Pop-III stellar mass of 8000~$M_\odot$, while the no LW case reaches high $\Delta\chi^2$ with 200~$M_\odot$.%\vid{For the top left Planck likelihood gray points, you basically use total tau prior from Planck? For the top right I can't see the crosses (for 10xMmin case), are they just hard to see? Bottom right the brown/green points don't have their no LW counter part? I like the plots I a lot, it really shows it is hard to get low total tau and low high-z tau tail with cmb EE data. Why are the delta chisquare colour-coded pieces having local maxima in most of the plots except last (bumps vs dips)?}
}
\label{fig:fidLW_dchi2_Sandro}
\end{figure*}

We now investigate whether future CMB surveys, which will push down the upper limit of $\tau(z>15)$ and measure the low-$\ell$ EE power spectrum at higher signal-to-noise, will help constrain Pop-III models.  The top-left panel of Figure~\ref{fig:fidLW_dchi2_Sandro} shows the $\Delta\chi^2$ values of the EE power as a function of total $\tau$ between each fiducial Pop-III model and the corresponding Pop-II-only model with the same total $\tau$.  We note that the $\Delta\chi^2$ values decreases by $\lesssim1$ if we choose to quote the minimum $\Delta\chi^2$ by looping over a series of Pop-II-only models spanning a large range of total $\tau$, which is what one would really have to do to claim a significant detection of Pop-III ionization.  The vertical black lines show the {\it Planck} 2-$\sigma$ constraint on the total $\tau$.  The points are color-coded by the values of $\tau(z>15)$, which are determined by the 6 values of the Pop-III SFE.  The maximum total $\tau$ value of the points is determined by the Pop-II escape fractions (see Table~\ref{tab:params}), while the minimum is set by our requirement that reionization completes before $z=5.5$.  For comparison, we also show the effective $\Delta\chi^2$ values using the {\it Planck} {\tt SimAll} likelihood in gray, defined as twice the difference between the log likelihoods of each Pop-III and Pop-II-only pair.\footnote{For the {\it Planck} effective $\Delta\chi^2$ values, looping over all Pop-II-only models and minimizing $|\Delta\chi^2|$ for a Pop-III model would always give 0, so we only choose to quote the effective $\Delta\chi^2$ compared at the same total $\tau$.}

A clear rising trend of $\Delta\chi^2$ with $\tau(z>15)$ implies that a $\tau(z>15)$ of at least $0.009$ is needed for a fiducial Pop-III model to be distinguishable at $>2$-$\sigma$ level from a Pop-II-only model with the same total $\tau$.\footnote{For fixed $\tau(z>15)$, $\Delta\chi^2$ between the Pop-III and Pop-II models first drops with increasing total $\tau$ as the differences in the EE power between the Pop-III model and the Pop-II-only model decrease at $\ell\lesssim10$, and then rises owing to the increasing difference in the EE power at $\ell\sim10-30$.  As total $\tau$ gets even higher, $\Delta\chi^2$ lowers again with reduced differences in the EE power at $\ell\sim10-30$.  Overall, the changes of $\Delta\chi^2$ with total $\tau$ when fixing $\tau(z>15)$ are less than $1.5$, within the range of parameter space that we have explored.}
To get to this level of $\tau(z>15)$, $\xion(z=15)$ has to reach $10-20\%$ (see Figure~\ref{fig:model_Sandro_master}).  Assuming that the Pop-III SFR density is roughly constant within a a recombination time of the IGM, the required SFR density to produce such level of $\xion$ can be estimated as
\begin{equation}
{\rm SFRD} = 3\times10^{-4} \left( \frac{\xion}{0.15} \right) \left( \frac{C}{3} \right) \left( \frac{N_{\rm ion}}{60,000} \right)^{-1} \left( \frac{1+z}{16} \right)^3 \ M_\odot/{\rm yr/Mpc^3},
\label{eq:sfr_estimate}
\end{equation}
where $N_{\rm ion}$ is number of ionizing photons produced per stellar baryon in Pop-III stars.  This analytic calculation appears to be consistent with our modeling results in Figure~\ref{fig:fidLW_Cl} and \ref{fig:model_Sandro_master} at factor of 2 level, although the non-constant SFR density in the fiducial model leads to potential errors of our estimates.

The effective $\Delta\chi^2$ values calculated using the {\it Planck} {\tt SimAll} likelihood are much lower than the cosmic variance limited $\Delta\chi^2$ values and are all less than 1 (see grey points in top-left panel of Figure~\ref{fig:fidLW_dchi2_Sandro}).  This result that the {\it Planck} data contains almost no information on $\tau(z>15)$ seems somewhat inconsistent with the revised $\tau(z>15)<0.018$ 2-$\sigma$ upper limit obtained by \citet{planck18, planck21correction}.
% Following the effective reionization likelihood approach of \citet{heinrich21}, we find that our reionization histories with $\tau(z>15)\gtrsim0.016$ produce $\Delta\chi^2>4$, roughly consistent with their analysis.
%Our results suggests that the current {\it Planck} data does not constrain an extended tail from Pop-III models.  
To understand this discrepancy, we attempted to replicate the revised constraint of \citet{planck18}, and we find that we can qualitatively reproduce their upper limit even if we assume the {\it Planck} data only has information on total $\tau$ and no other property of the ionization history.  In particular, we partially mimic the \citet{planck18} analysis by sampling the likelihood for $20,000$ FlexKnot reionization histories with 5 knots and selecting histories such that they produce a uniform distribution of $\tau(z>15)$ in $[0,0.04]$.  To define a likelihood of a reionization history that only depends on the total $\tau$, we take the likelihood of a reionization history to be the same as the {\tt SimAll} likelihood of a single tanh with the same total $\tau$.  We find that this approach gives a 2-$\sigma$ upper limit of $\tau(z>15)=0.015$, close to the values obtained by the revised {\it Planck} and \citet{heinrich21}.  Our findings suggest that the \citet{planck18} inferred $\tau(z>15)$ upper limit may simply reflect the $\tau(z>15)$ values from the set of FlexKnot reionization histories that are consistent with the total $\tau$ constraints.  We have not investigated whether a similar conclusion applies to the principal component analysis of \citet{heinrich21}, which find a similar $2\sigma$ constrain to \citet{planck18} of $\tau(z>15)<0.020$.\footnote{We note that our procedure does not fully follow that of \citet{planck18}, which marginalizes over the number of knots, weighting by Bayesian evidence.  In the  \citet{planck18} analysis, the $\tau(z>15)$ posterior is shaped primarily by lower knot models than 5-knots, models for which we expect total $\tau$ to be more correlated with $\tau(z>15)$.  Our uniform $\tau(z>15)$ distribution approximates \citet{planck18}'s flat prior on $\tau(z>15)$.}
% \tmp{We also note that the Fisher matrix analysis of \citet{watts20} showed that a 3-$\sigma$ detection of $\tau(z>15)$ at the {\it Planck} noise level should correspond to a value of $0.024$, so any model that produces a $\tau(z>15)$ lower than this value is not expected to be distinguishable from other models with the same total $\tau$.  This further confirms our suspicion that the {\it Planck} data and likelihood are unlikely to significantly constrain $\tau(z>15)$.  Our prediction on the required $\tau(z>15)$ to obtain $>2$-$\sigma$ detection of an extended reionization tail for future CMB surveys is thus not in tension with the \citet{planck18} and \citet{heinrich21} $\tau(z>15)$ constraints, which we suspect are mostly bounding the $\tau(z>15)$ range that are consistent with the {\it Planck} total $\tau$.}

Since \citet{planck18} and \citet{heinrich21} suggest that reionization histories consistent with the {\it Planck} total $\tau$ are allowed to have $\tau(z>15)\lesssim0.020$, the fact that our models do not support such high $\tau(z>15)$ when producing low total $\tau$ may seem in tension with their results.  The main reason why our models do not allow high $\tau(z>15)$ when being consistent with the {\it Planck} total $\tau$ is that is that our when our fiducial generates high $\tau(z>15)$, Pop-III stars do not quench rapidly at $z<15$ but continues to contribute significantly to reionization, leading to our models reaching 50\% ionization too early.  We discuss possible models that create flatter $z>15$ ionization or ``double reionization'' which gets around this issue in Section~\ref{sec:exotic_models}.  Having a faster Pop-II reionization than the \citet{tacchella18} model is unlikely to help allow more $\tau(z>15)$ from Pop-III, since a fixed Pop-III star formation efficiency produces $<0.002$ differences in $\tau(z>15)$ when coupled to the much more extended \citet{furlanetto17} Pop-II model (bottom left panel of Figure~\ref{fig:fidLW_dchi2_Sandro}).  Moreover, as we have pointed out in Section~\ref{sec:methods}, the rapid star formation history of \citet{tacchella18} is already in tension with Lyman-$\alpha$ forest observations, which favors a much flatter emissivity history.
We also point out that the FlexKnot model creates large fluctuations in the reionization histories and the first two principle components of \citet{heinrich21} carry a lot of contribution to $\tau(z>15)$, while our physically motivated models do not produce big peaks and troughs in the reionization history at $z>15$.  It is unclear whether any physically motivated models can creates these types of reionization histories.

To check the robustness of the above $\tau(z>15)>0.009$ threshold for obtaining $\Delta\chi^2>4$, we examine the effects of varying the strength of LW feedback and the Pop-II SFR.  The top-right panel of Figure~\ref{fig:fidLW_dchi2_Sandro} presents models with no LW (triangles), fiducial $M_{\rm min}$ (dots), and $10 \times M_{\rm min}$ (crosses, piled at the bottom of the plot), where in the no LW case we adjusted the SFE so that the models produce roughly the same set of $\tau(z>15)$ values as the fiducial case.  The $\Delta\chi^2$ values among different models are consistent with each other when fixing $\tau(z>15)$, owing to similar shapes of the Pop-III SFR.
We note that the no LW model requires much lower SFE in order to produce the same $\tau(z>15)$ as the fiducial model, and a SFE of $3\times10^{-4}$ already yields total $\tau$ higher than the 2-$\sigma$ upper limit by {\it Planck}.  This indicates that the total $\tau$ constraint rules out models without negative feedback \citep[see also][]{haiman06}.
The bottom-left panel compares models using the fiducial \citet{tacchella18} Pop-II model (dots) against the more extended \citet{furlanetto17} model (stars).  The much larger contribution of Pop-II stars to ionization at $z>15$ in the \citet{furlanetto17} model boosts the $\tau(z>15)$ threshold for obtaining $\Delta\chi^2>4$ to $0.014$, making it even more difficult for Pop-III models to be distinguishable from Pop-II-only ones using the low-$\ell$ EE power spectrum.  The $\tau(z>15)$ threshold to get $\Delta\chi^2>4$ is thus model-dependent.

The $\Delta\chi^2$ values of the fixed-stellar-mass model, explored by the bottom-right panel of Figure~\ref{fig:fidLW_dchi2_Sandro}, appears to be even smaller when using the fiducial LW feedback (dots) owing to much lower Pop-III SFR produced by such models compared to the fiducial ``\dfcolldt'' model.  Unless the Pop-III stellar mass can be as unphysically high as $8000\ M_\odot$, $\Delta\chi^2$ remains $\ll4$.  Assuming no LW (triangles), on the other hand, produces $\Delta\chi^2 > 4$ at a Pop-III stellar mass of $100\ M_\odot$.  This owes to the much flatter Pop-III SFR produced by the fixed-stellar-mass model, and implies the possibility of getting a more distinguishable Pop-III model by using the fixed-stellar-mass model with weak LW feedback, which we will explore in Section~\ref{sec:exotic_models}.
However, the no LW models with $\ge100\ M_\odot$ also easily overshoot the total $\tau$ outside the 2-$\sigma$ constraint by {\it Planck}, while a Pop-III stellar mass of a few hundred $M_\odot$ per halo seems more consistent with recent simulations \citep[e.g.][]{susa14, hirano15, stacy16}.  This echos our previous statement regarding the ``\dfcolldt'' model that the total $\tau$ constraint rules out models without negative feedback.

In summary, neither the fiducial ``\dfcolldt'' model nor the fixed-stellar-mass model with the \citet{machacek01} LW feedback is able to produce high enough $\tau(z>15)$ so that the model can be distinguished at $>2$-$\sigma$ level from a Pop-II-only model with the same total $\tau$ using the low-$\ell$ EE power spectrum, while keeping the total $\tau$ consistent with the {\it Planck} constraints.  In order for a Pop-III model to produce high $\tau(z<15)$ and low total $\tau$ so as to make the CMB a useful tool for constraining Pop-III models, the Pop-III star formation history has to be either more plateaued or quenched at $z\sim10-15$, so that when Pop-II starts dominating reionization at lower redshifts, the ionizing photon budget from Pop-III hardly contribute to increasing the total $\tau$ anymore.  We explore the possible forms of models capable of producing such Pop-III star formation histories in Section~\ref{sec:exotic_models}.

\section{Pop-III models that can leave detectable imprints on the CMB}
\label{sec:exotic_models}

\begin{figure*}
\includegraphics[width=2\columnwidth]{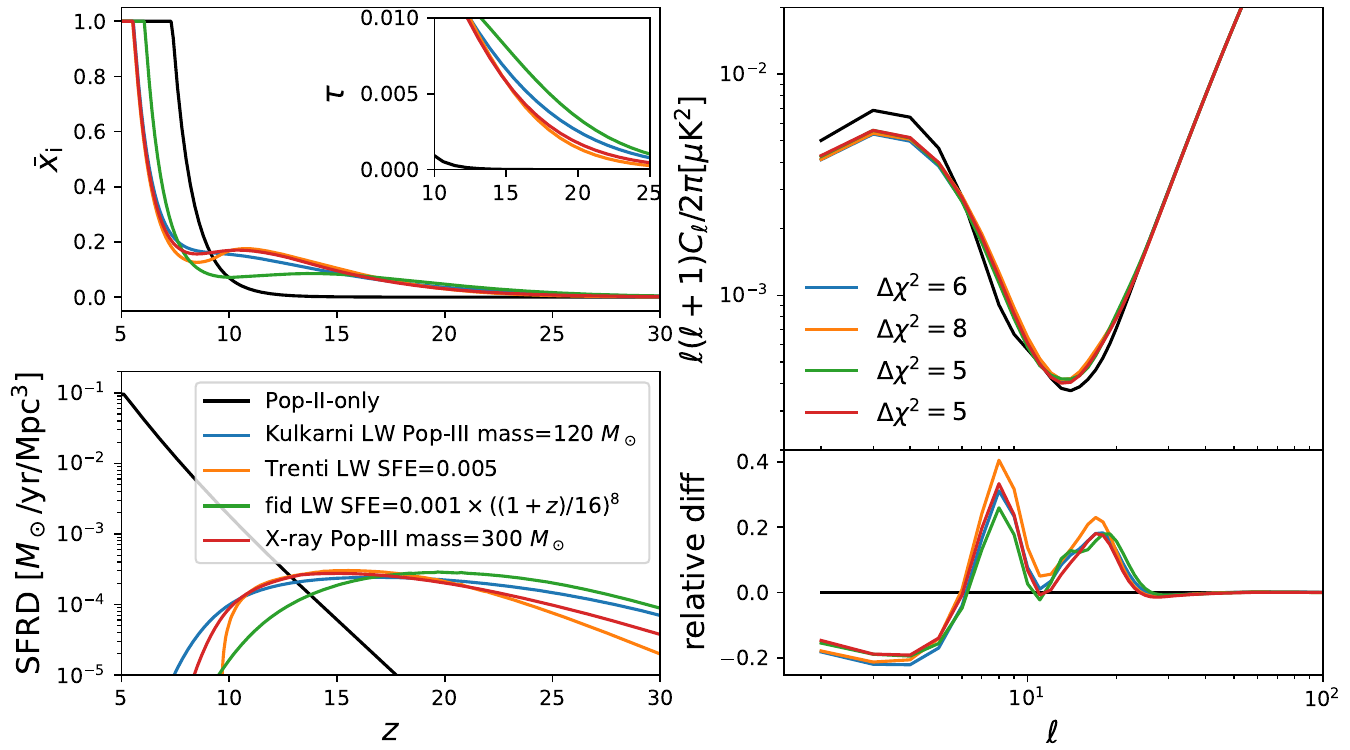}
\caption{Possible forms of models that produce $\Delta\chi^2\ge4$ for a cosmic variance-limited EE power spectrum measurement.  All models have the same total $\tau=0.060$.  Black: The fiducial Pop-II-only model, which is the reference model for the $\Delta\chi^2$.  Blue: the fixed-stellar-mass model with the \citet{kulkarni20} LW feedback and 120~$M_\odot$ stellar mass per halo, giving $\tau(z>15)=0.0066$.  Orange: the fiducial ``\dfcolldt'' model with a modified version of the \citet{trenti09} LW feedback, producing $\tau(z>15)=0.006$.  Green: the fiducial ``\dfcolldt'' model with \citet{machacek01} LW feedback and a Pop-III SFE that goes as $0.001((1+z)/16)^8$, resulting in $\tau(z>15)=0.008$.  Red: the fixed-stellar-mass model with a framework mimicking the effects of X-ray feedback.  While all of these models can produce a detectable signature, we argue in \S~\ref{sec:exotic_models} that most of the models must make exotic and tuned assumptions.   The top-left panel shows the reionization histories, and the inset illustrate the accumulated $\tau$ at $z=10-25$.  bottom-left panel presents the SFR densities.  The right panels show the EE power spectra and the $\Delta\chi^2$ values (top panel), and the relative differences of the Pop-III models to the Pop-II-only one (bottom panel).}
\label{fig:models_max_chi2}
\end{figure*}

In Section~\ref{sec:main_results} we showed that the fiducial model fails to produce high $\tau(z>15)$ and low total $\tau$ at the same time.  Here we discuss possible forms of Pop-III models that have a prolonged plateau of Pop-III star formation or have it quenched at $z\sim10-15$.  Figure~\ref{fig:models_max_chi2} compares four types of models considered below, with the parameters tuned so that the models produce $\Delta\chi^2>4$ and low total $\tau=0.060$ consistent with \citet{planck18}.  Black lines show a Pop-II-only model with total $\tau=0.060$.  Top-left panel shows the reionization histories, and the inset illustrate the accumulated $\tau$ at $z=10-25$.  Bottom-left panel presents the SFR densities.  The right panels illustrate the EE power spectra (top panel) and the relative differences of the Pop-III models to the Pop-II-only one (bottom panel).  All Pop-III models result in $\Delta\chi^2>4$ of the EE power spectra at $\ell=2-100$ as indicated in the top-right panel, thus detectable at $\sqrt{\Delta\chi}$ standard deviations with future CMB surveys.  The models also have Pop-III SFR densities of $\sim2\times10^{-4}\ M_\odot/$yr/Mpc$^3$ at $z=15$, roughly consistent with our estimates surrounding equation~\ref{eq:sfr_estimate} for what is required to get a signal detectable at $\gtrsim 2\sigma$.
We note that we choose to let all models to have the same ({\it Planck}-consistent) total $\tau=0.060$ simply for the convenience of calibrating parameters.  For the models shown in Section~\ref{sec:exotic_kulkarni} and \ref{sec:exotic_xray}, allowing higher Pop-III stellar mass per halo can give higher $\tau(z>15)$ and $\Delta\chi^2\sim9$, without exceeding the {\it Planck} total $\tau$ constraint by 2-$\sigma$.
%While a time-varying clumping factor may also produce plateaued reionization or ``double reionization'' features, we only focus on aspects in Pop-III star formation and keep the clumping factor constant.  %We note that a clumping factor that grows rapidly with decreasing redshift can also help flatten the reionization history at $z\gtrsim15$, but the effects of a redshift-dependent clumping factor can be degenerate with the effects of a redshift-dependent star formation efficiency.  Furthermore, as we argued in Section~\ref{sec:methods}, previous works have debated extensively on the value and evolution of the clumping factor at $z\gtrsim6$.  We therefore refrain from exploring the possibility of time-varying clumping factors without the aid of detailed numerical simulations, but only discuss possible forms of Pop-III star formation models capable of producing the desired types of Pop-III star formation histories under the assumption of a constant clumping factor.

\subsection{Models using the fixed-stellar-mass star formation with a weaker LW feedback}
\label{sec:exotic_kulkarni}

This type of model is motivated by the fixed-stellar-mass model being able to produce $\Delta\chi^2>4$ and low total $\tau$ with physically motivated values of the Pop-III stellar mass owing to the much flatter Pop-III SFR than the fiducial ``\dfcolldt'' model (see the bottom-right panel of Figure~\ref{fig:fidLW_dchi2_Sandro}).  We therefore test the $M_{\rm min}$ prescription given in \citet{kulkarni20}, who found much weaker effects of LW feedback than previous works owing to gas self-shielding.  For $J_{\rm LW}=0.01-1 \times 10^{-21}$~erg/s/cm$^2$/Hz/Sr, the \citet{kulkarni20} $M_{\rm min}$ is a factor of $10-20$ smaller than the canonical \citet{machacek01} $M_{\rm min}$.\footnote{Because of the much weaker feedback and because it assumes a fixed stellar mass, this \citet{kulkarni20} model is somewhat similar to the fixed stellar mass, no Lyman-Werner feedback models in Fig.~\ref{fig:fidLW_dchi2_Sandro}, which yielded our largest $\Delta \chi^2$ of the models we considered there.}

The blue lines in Figure~\ref{fig:models_max_chi2} represent this \citet{kulkarni20} model assuming a Pop-III stellar mass of 120~$M_\odot$, which gives $\tau(z>15)=0.0066$.  The model behaves similarly to a no-LW model where Pop-III stars form in all halos with $T_{\rm vir}\ge500$~K owing to the very low amplitude of $M_{\rm min}$ being only a factor of $1.5$ higher than $T_{\rm vir}=500$~K.  The much flatter shape of the Pop-III SFR in the fixed-stellar-mass model allows significant Pop-III reionization at $z>15$ under such weak LW feedback, while in the mean time limiting the contribution of Pop-III stars to reionization at later redshifts compared to the \citet{machacek01} model.
120~$M_\odot$ is roughly the minimum Pop-III stellar mass per halo to get $\Delta\chi^2>4$.  Higher Pop-III stellar masses raise $\Delta\chi^2$, and up to 200~$M_\odot$ can all produce total $\tau$ consistent with the Planck constraint within 2-$\sigma$.

While both the fixed-stellar-mass model and weak LW feedback seem consistent with recent simulation works, the very low amplitude of $M_{\rm min}$ in the \citet{kulkarni20} model has not been verified by other simulations.  For instance, \citet{schauer20, skinner20} only find a factor of two smaller $M_{\rm min}$ in their simulations than \citet{machacek01}.  We find that a factor of 2(5) decrease in the amplitude of the fiducial \citet{machacek01} $M_{\rm min}$ requires a Pop-III stellar mass per halo of 4000(800)~$M_\odot$ to get $\Delta\chi^2>4$, which is too high compared to the a few hundred $M_\odot$ Pop-III mass per halo found by many simulations \citep[e.g.][]{susa14, stacy16, park21a}.\footnote{\citet{kulkarni20} also finds that baryon-dark matter streaming results in a large suppression that increases with increasing redshift, which suggests that our implementation of their model likely results in a more plateaued history than they would expect.  This baryon streaming effect, where there is some controversy over how important it is, always goes in the direction of suppressing star formation at high redshifts \citep{oleary12, kulkarni20, schauer20}. Accounting for it in our models would strengthen our conclusions.}

\subsection{Models using the ``\dfcolldt'' star formation with a stronger LW feedback}
\label{sec:exotic_trenti}

Specifically, we explore the functional form of $M_{\rm min}$ presented in \citet{trenti09} which has a steeper $(1+z)^{-3.6}$ scaling with redshift, while the fiducial \citet{machacek01} model has a $(1+z)^{-1.5}$ scaling.  This functional form comes from an analytic analysis comparing the H$_2$ formation time, photodissociation time, and the Hubble time, although it does not appear to be reproduced in simulations and so we consider it an extreme model.  We find that the original $M_{\rm min}$ formula in \citet{trenti09} results in too low Pop-III SFR densities that only reach $\sim10^{-5}\ M_\odot$/yr/Mpc$^3$ and hence are not constrained by CMB measurements (see Appendix~\ref{sec:appendix_LW}).  To fix this, we  lowered the amplitude of the original $M_{\rm min}$ by a factor of three while keeping the original $(1+z)^{-3.6}$ scaling of $M_{\rm min}$. % which can be justified by self-shielding of the halo gas.

This model is shown by the orange lines in Figure~\ref{fig:models_max_chi2}, where we adopt a star formation efficiency of $0.005$, resulting in $\tau(z>15)=0.006$.  With more aggressive LW feedback at lower $z$, the \citet{trenti09} functional form of LW feedback produces more flattened Pop-III star formation histories than the \citet{machacek01} LW feedback for the ``\dfcolldt'' model at $z\sim10-20$ when Pop-III star formation begins to saturate, and it eventually quenches Pop-III stars at $z<10$.  The \citet{machacek01} model, on the other hand, allows Pop-III star formation until the end of reionization (e.g. Figure~\ref{fig:fidLW_Cl}).  The \citet{trenti09} form of LW feedback thus creates a ``double reionization'' type of feature, leaving the most distinct imprint of any of the explored models with $\Delta \chi^2 = 8$ on the large-scale CMB E-mode polarization.  We note that the adopted SFE of $0.005$ is very high compared to the more consensus value $0.001$ used in the literature.  To be able to reach the same $\tau(z>15)$ with lower SFE requires lowering $M_{\rm min}$, but the saturation of the Pop-III SFR is lost when using weaker LW feedback strength.  It is thus very hard to achieve low total $\tau$ and high $\tau(z>15)$ at the same time using a more physically plausible SFE with lower $M_{\rm min}$.  Moreover, the \citet{trenti09} $M_{\rm min}$ has not be found in numerical simulations and recent works also suggest lower $M_{\rm min}$ owing to gas self-shielding \citep{schauer20, kulkarni20, skinner20}.

\subsection{Models with star formation efficiencies that increase with redshift}
\label{sec:exotic_SFE}

Here we try to understand how strongly we need to alter the scaling of the star formation efficiency with redshift (or, almost equivalently, the inverse of the clumping factor) in our fiducial models.  Thus, we will only consider the fiducial ``\dfcolldt'' plus \citet{machacek01} LW model, and we include an ad hoc star formation efficiency boost factor of $(1+z)^n$, where $n$ is a free parameter that is tuned to yield significant $\Delta \chi^2$.  We fix the star formation efficiency at $z=15$ to be $0.001$, which is an arbitrary choice but implemented to reduce the degrees of freedom.  %The choice of this fixed point is arbitrary but the main purpose is to limit the degree of freedom in varying this family model.
Perhaps a motivation for this ad hoc star formation efficiency boost could be that, at fixed mass, the gravitational potential well of halos being deeper at higher redshifts, or Pop-III star formation is quenched by metal enrichment at lower $z$, although such a motivation would likely be insufficient for large values of $n$.

Our results in Section~\ref{sec:main_results} have shown that for the ``\dfcolldt'' model with the \citet{machacek01} LW feedback, a constant Pop-III star formation efficiency of at least $0.01$ is needed to produce high enough $\tau(z>15)$ to get $\Delta\chi^2>4$.  Since we fix the Pop-III star formation efficiency at $z=15$ to be $0.001$, the star formation efficiency has to quickly reach a mammoth value of $\sim0.01$ at $z=15-20$ in order to produce high $\tau(z>15)$.  On the other hand, the star formation efficiency needs to drop quickly to $\lesssim10^{-4}$ at $z\lesssim15$ so that Pop-III stars no longer contribute significantly to the total $\tau$ at lower $z$.  These considerations resulted in us finding that we need an extremely strong $(1+z)^8$ scaling of the boost factor in order to produce $\Delta\chi>4$ and low total $\tau$.  Allowing for both a higher total $\tau < 0.08$ and a fixed point at lower $z$ can mitigate such strong scaling, but may still require a steep scaling as $(1+z)^4$ to obtain $\Delta\chi^2>4$.  The green lines in Figure~\ref{fig:models_max_chi2} illustrates the model with the $(1+z)^8$ scaling, leading to $\tau(z>15)=0.008$ and $\Delta\chi=5$.  We cannot motivate such a strong boost of the SFE with redshift.  If we instead altered the clumping factor rather than the SFE, this experiment argues that the clumping factor would have to scale as $(1+z)^{-8}$ to flatten the reionization history at $z\gtrsim15$ (c.f.~eqn.~\ref{eq:sfr_estimate}), a trend that is opposite to our expectation (\S~\ref{sec:methods}).

\subsection{Models with X-ray feedback}
\label{sec:exotic_xray}

The effect of X-ray feedback on Pop-III star formation can be positive at higher $z$, while negative at later times.  An X-ray background can be created by the supernova/hypernova explosion of Pop-III stars, accretion onto intermediate mass black holes from Pop-III, and Pop-III X-ray binaries \citep{jeon14xray, xu16, ricotti16}.  It can partially ionized the IGM, thus increasing H$_2$ formation through H + H$^- \rightarrow$ H$_2$ + e$^-$.  Meanwhile, X-ray heating raises the gas temperature and lifts the IGM Jeans mass, counteracting the positive feedback on H$_2$ cooling \citep{haiman00, venkatesan01, oh01}.  It is thus likely that X-rays boost Pop-III star formation at early times and suppress it at late times owing to the increased heating \citep{ricotti16}, producing the desired prolonged reionization or ``double reionization'' feature that can be detectable using the low-$\ell$ CMB.

Despite being prospective of creating detectable imprints on the CMB, the effects of X-ray feedback have been greatly debated in the literature.  While most studies find that X-rays likely ionize the IGM to a few percent level and heat it to about 1000~K \citep[e.g.][]{kuhlen05, xu16}, certain prescriptions for X-ray emission can lead to 10-20\% ionization of the IGM and a temperature of over $10^4$~K at $z\gtrsim15$ \citep[e.g.][]{venkatesan01, ricotti04, jeon14xray}.  The impact on Pop-III star formation is also an open question.  Some studies found no strong net feedback of X-rays on the Pop-III star formation history owing to the cancellation of the increased H$_2$ cooling and the heating \citep[e.g.][]{machacek03, kuhlen05, jeon14xray} or X-ray self-shielding at high gas densities \citep{hummel15}, while others demonstrated a factor of 10 increase in the Pop-III star-forming halos \citep{jeon12, park21a}, although with a factor of $\sim2$ decrease in the Pop-III multiplicity and stellar mass as well \citep{park21b}.  However, studies have also shown that X-rays can reduce the IGM clumping at a factor of $\sim3$ level \citep[e.g.][]{kuhlen05, jeon14xray}, resulting in a positive feedback on the reionization process.

Given the above uncertainties, we do not attempt to model the effects of X-ray feedback self-consistently.  However, we note that for moderate X-ray intensities, the semi-analytic modeling of \citet{ricotti16} predicts that the change in $M_{\rm min}$ in the presence of X-ray feedback is mostly a decrease in the amplitude of $M_{\rm min}$, except at later times when the increased X-ray heating gradually quenches Pop-III star formation.  We thus mimic the effects of early-time positive X-ray feedback by lowering $M_{\rm min}$ of the \citet{machacek01} model by a factor of 10 \citep{ricotti16, park21a}, while at late times determine $M_{\rm min}$ via $T_{\rm vir}>T_{\rm IGM}$, where the IGM temperature $T_{\rm IGM}$ is converted from the LW background intensity using equation~13 of \citet{ricotti16}.\footnote{We note that since we do not calculate X-ray ionization of the IGM self-consistently, we take the $f_h$ factor in \citet{ricotti16} to be a constant value of $0.26$, corresponding to an electron fraction of 1\%.  However, $f_h$ changes by a factor of 4 from no ionization to 10\% electron fraction.}  We couple this framework of X-ray feedback with the fixed-stellar-mass model.  We do not include possible changes in the clumping factor owing to X-ray heating of the IGM, but we note that \citet{jeon14xray} found a flat $C$ with time, consistent with our choice of constant $C$.

The red lines in Figure~\ref{fig:models_max_chi2} show such a model mimicking X-ray feedback, assuming a Pop-III mass per halo of $300\ M_\odot$, and that the mean energy emitted in X-ray bands is 1\% of that emitted in LW bands.  This 1\% ratio is in between the X-ray energy released by supernova and hypernova \citep[$\beta_X=0.01$ in][]{ricotti16}.  The negative feedback of X-ray heating starts dominating the control of $M_{\rm min}$ at $z=13$, leading to a ``double reionization'' feature which results in $\tau(z>15)=0.006$ and $\Delta\chi^2=5$ of the EE power spectrum.  Using a Pop-III stellar mass per halo of $500\ M_\odot$ gives total $\tau=0.067$, $\tau(z>15)=0.007$, and $\Delta\chi^2=8$.

If X-ray feedback operates in a manner like how it has been tuned above, strong X-ray feedback would allow Pop-III reionization to have a $3\sigma$ detectable effect in a cosmic variance-limited CMB observation.\footnote{Studies have also shown that X-rays can pre-ionize the IGM at percent level \citep[e.g.][]{xu14}.  A 1\% ionization from $z=15$ to 30 would contribute an optical depth of just $0.002$.  This scenario may be only marginally detectable in future CMB surveys \citep{watts20}.  Realistically, X-ray ionization would have to reach extreme values of 10\% at $z\sim 15$ to leave detectable imprints on the CMB if the ionization traces the Pop-III star formation in our models (as few stars form at $z>20$).  The lower recombination rate in the partially X-ray--ionized gas makes it more difficult to have as plateaued a history in this scenario as the UV ionization scenarios we have considered.}
However, a full exploration of X-ray feedback is needed to confirm this and is beyond the scope of our paper.  We also note that there are a number of other possible effects that might change the shape of the Pop-III star formation history or reionization history.  These include a possible suppression of Pop-III star formation owing to the relative velocity between baryons and dark matter \citep{kulkarni20, schauer20}, quenching of Pop-III star formation owing to metal enrichment \citep{trenti09b, visbal20}, positive UV feedback on Pop-III star formation \citep{ricotti01, ricotti02}, and the ionized bubbles created by supernova explosion \citep{abe21planck}.  Incorporating all of these self-consistently into modeling Pop-III star formation is challenging.  Capturing LW feedback, which is generally believed to be the most important feedback mechanisms shaping Pop-III star formation, is more crucial for producing the simplest possible physically motivated reionization histories in our work.\\

In summary, the required fine-tuning of the parameters in most of the above exotic models or the lack of support of such models from other simulations echoes our conclusion in Section~\ref{sec:main_results} that the CMB is unlikely to be a useful constraint for Pop-III models.

\section{Conclusions}
\label{sec:conclusions}

We have examined the constraining power of current and future large-scale CMB E-mode polarization observations using semi-analytic PopII + Pop-III star formation models.  We calculated the reionization histories for a variety of semi-analytic Pop-III models, with the fiducial model assuming that the Pop-III star formation rate is proportional to the star formation efficiency times halo accretion rate (the ``\dfcolldt'' model) and is determined by the \citet{machacek01} type of LW feedback, and that Pop-II star formation follows the model of \citet{tacchella18}.  We found that  $\tau(z>15)>0.009$ is required for the fiducial model to be distinguished from a Pop-II-only model at $>2$-$\sigma$ level for a cosmic variance-limited measurement.  However, such models produce too high total $\tau$ that is inconsistent with \citet{planck18}.  This $\tau(z>15)$ threshold is robust with changes in the strength of the LW feedback, and is boosted even higher if Pop-II star formation is more extended at higher redshifts.  Compared to the fiducial model, models assuming a fixed amount of Pop-III stellar mass forming in each halo have flatter Pop-III star formation histories, but are unable to produce high enough $\tau(z>15)$ for reasonable values of the Pop-III stellar mass, unless LW feedback is very weak.  Our modeling results imply that future observations of the CMB EE power spectrum at $\ell=10-20$ and $\tau(z>15)$ are unlikely to put further constraints on Pop-III models than the total $\tau$ and endpoint of reionization constraint as imposed by the Lyman-$\alpha$ forest.
While recent studies find the posterior peaks at total $\tau$ that are shifted up by half a standard deviation using the {\it Planck} data \citep{pagano20, debel21, beyondplanck20}, our main conclusions only significantly change if $\tau$ were a few sigma larger than found by \citet{planck18}. 

Since standard Pop-III models are unlikely to be distinguished from Pop-II-only models using the large-scale CMB EE power spectrum at $>2$-$\sigma$ level, we explored more exotic types of models that can be more distinguishable using the CMB.  We find that models where Lyman-Werner feedback scales strongly with redshift or Pop-III star formation efficiency changes rapidly with redshift are more likely to produce a sufficiently plateaued ionization to be detectable at $2\sigma$.  Such a plateau also likely occurs in models assuming a fixed stellar mass forms per halo and a considerably weak Lyman-Werner feedback.
However, none of these models are supported by other semi-analytic works or numerical simulations.  Even with these types of models, adjusting the star formation efficiency or Lyman-Werner feedback strength so that the models can produce high $\tau(z>15)$ and low total $\tau$ at the same time could be hard to achieve.  This further demonstrate the difficulty of using the CMB to constrain Pop-III models.
However, existing large modeling uncertainties allow for X-ray feedback to be tuned in a manner to turn off Pop-III star formation at late times, plateauing the ionization sufficiently to leave a potentially detectable signature. We leave it to future work to evaluate whether our imagined strong X-ray feedback scenario is achievable.

While we find it is difficult to have a detectable effect on the CMB, another interesting question is whether an extended reionization could bias precision cosmological estimates using the standard tanh-type reionization analysis.  Our $\Delta \chi^2$ values between our most extended reionization histories and a tanh-like model reach $\sim 5$. Further, the E-mode power differences do not look particularly like another parameter, with the parameters of most relevance likely being `lever-arm' parameters like neutrino mass and scalar spectral index.  While we suspect the bias our histories would impart for standard cosmological parameters is small, this is an interesting question for future work to address.  The parameter biases of our physical reionization history models are certain to be smaller than the flexible principal component or FlexKnot models.

Finally, we note that if we calculate effective $\Delta\chi^2$ values of the EE power spectra using the {\it Planck} {\tt SimAll} likelihood, models with $\tau(z>15)$ as large as $0.015$ only have effective $\Delta\chi^2 < 1$ when compared to models with the same total $\tau$ but negligible $\tau(z>15)$.  Thus, for the physical reionization histories we have explored, the {\it Planck} likelihood is mostly sensitive to the total $\tau$ and provides little independent constraint on $\tau(z>15)$. Such small $\Delta\chi^2$ seem somewhat inconsistent with the recently revised results of \citet{planck18} that $\tau(z>15) <0.018$ at 2-$\sigma$ level and similarly the $\tau(z>15) <0.020$ of \citet{heinrich21}.  We briefly discussed how the \citet{planck18} $\tau(z>15)$ constraints may be driven by the \citet{planck18} total $\tau$ and not reflect significant additional information on $\tau(z>15)$ in the {\it Planck} likelihood.

\section*{Acknowledgements}

We thank Sandro Tacchella, Rick Mebane, Jordan Mirocha, Julian Munoz, Duncan Watts, Zoltan Haiman, and Chen Heinrich for useful conversations during this work.  We especially thank Marius Millea for providing valuable information regarding the {\it Planck} analysis.
MM is supported by NASA award 19-ATP19-0191.  DJE is partially funded by NASA contract NAS5-02015 and as a Simons Foundation Investigator. VI is supported by the Kavli foundation.

\section*{Data Availability}

The data underlying this article will be shared on reasonable request to the corresponding author.

\bibliographystyle{mnras}
\bibliography{example}

\appendix

\section{Modeling the ionization history}
\label{sec:appendix_model}

While we calculate the ionization history using equation~\ref{eq:dxidt}, a self-consistent derivation of the time evolution of $\xion$ results in a quadratic term in $\xion$ \citep{chen20}:
\begin{equation}
\frac{\mathrm{d} \xion}{\dt} = \epsilon - C_R \nH \langle \alpha(T) \rangle_V \left( 1 + \frac{Y}{4X} \right) \xion^2,
\end{equation}
where $\langle \cdot \rangle_V$ denotes volume average, and the clumping factor $C_R$ is defined as
\begin{equation}
C_R = \frac{\langle \alpha(T) n_{\rm HII} n_{\rm e} \rangle_V}{\langle \alpha(T) \rangle_V \langle n_{\rm HII} \rangle_V \langle n_{\rm e} \rangle_V}.
\end{equation}
Here $n_{\rm HII}$ and $n_{\rm e}$ are the number densities of HII and electron respectively.  Owing to the lack of constrains on $C_R$, instead of modeling the quadratic form of the time evolution of $\xion$, we define $C \equiv C_R \xion$ which indicates the recombination rate in units of the recombination rate of a homogeneous universe that has the same ionization state \citep{finlator12} and take $\langle \alpha(T) \rangle_V = \alpha(T_0) = 2.6\times10^{-13}$~cm$^{-3}$s$^{-1}$.  This effectively expresses the recombination term as $\xion/\bar{t}_{\rm rec}$, where
\begin{equation}
\bar{t}_{\rm rec} = \frac{1}{(1+Y/4X)\alpha(T_0) C \nH}.
\end{equation}

\begin{figure}
\centering
\includegraphics[width=\columnwidth]{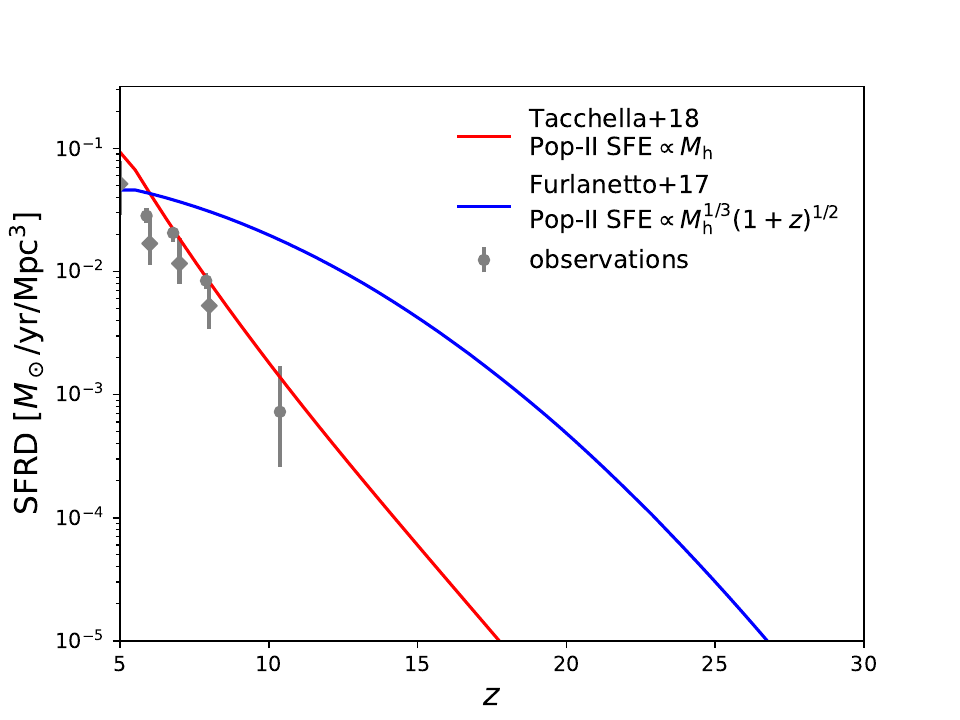}
\caption{SFR density as a function of redshift for the two Pop-II models used in this work.  Gray circles with errorbars represent the observations of \citet{bouwens15} and \citet{finkelstein15}.}
\label{fig:popIIsfrd}
\end{figure}

The ionizing photon emissivity is given by
\begin{equation}
\epsilon = \frac{{\rm SFRD_{III}}}{\rho_{\rm b}} f_{\rm esc,III} N_{\rm ion,III} + \frac{{\rm SFRD_{II}}}{\rho_{\rm b}} f_{\rm esc,II} N_{\rm ion,II},
\label{eq:dngammadt}
\end{equation}
where $\rho_{\rm b}$ is the mean baryon density, ${\rm SFRD_{III}},\ {\rm SFRD_{II}}$ are the star formation rate densities of Pop-III and Pop-II stars respectively, $N_{\rm ion,*}$ is the number of ionizing photons produced per stellar baryon, and $f_{\rm esc,*}$ is the escape fraction of ionizing photons of the star-forming halos.  As mentioned in Section~\ref{sec:methods}, we take $f_{\rm esc,III}=1, N_{\rm ion,III}=60,000,\ N_{\rm ion,II}=4000$, and vary $f_{\rm esc,II}$ as a free parameter.  ${\rm SFRD_{II}}$ is given by
\begin{equation}
{\rm SFRD_{II}} \propto f_{\rm II} \frac{\mathrm{d}\fcoll}{\dt} = f_{\rm II} \frac{\mathrm{d}}{\dt} \frac{1}{\Omega_m \rho_c} \int_{M_{\rm a}}^\infty \mathrm{d}M M \frac{\mathrm{d}n}{\mathrm{d}M},
\end{equation}
where $f_{\rm II}$ is the Pop-II star formation efficiency, $\fcoll$ is the cosmic mass fraction collapsed into dark matter halos with mass $M > M_{\rm a}$,  $\rho_c$ is the cosmological critical density, and $\mathrm{d}n/\mathrm{d}M$ is the Sheth-Tormen halo mass function.  The \citet{tacchella18} model and the \citet{furlanetto17} model take $f_{\rm II} \propto M$ and $f_{\rm II} \propto M^{1/3} (1+z)^{1/2}$ respectively, whose resulting Pop-II SFR densities are shown by the red and blue lines in Figure~\ref{fig:popIIsfrd} respectively, against the observations of \citet{bouwens15, finkelstein15} (grey points).  We have adjusted the $f_{\rm II}$ values from their original ones to fit our use of the Sheth-Tormen mass function.  The \citet{tacchella18} model gives good agreement of the Pop-II SFR density with the observations, while the \citet{furlanetto17} model overshoots the observational data points at $z>5$.  We note that faint galaxies in the \citet{furlanetto17} model contribute a significant fraction of the total Pop-II SFR density.  Including only the bright galaxies in the SFR density calculation would result in good agreement with the observations \citep{furlanetto17}.

Our fiducial model assumes that the Pop-III star formation rate density is determined by the star formation efficiency $f_{\rm III}$ and the dark matter accretion rate (which we term the ``\dfcolldt'' model):
\begin{equation}
{\rm SFRD_{III}} \propto f_{\rm III}\frac{\mathrm{d}\fcoll}{\dt} = f_{\rm III}\frac{\mathrm{d}}{\dt}\frac{1}{\Omega_m \rho_c} \int_{M_{\rm min}}^{M_{\rm a}} \mathrm{d}M M \frac{\mathrm{d}n}{\mathrm{d}M}.
\end{equation}
The fixed-stellar-mass model assumes that a fixed Pop-III stellar mass $M_{\rm III}$ can form in each newly collapsed halo, yielding
\begin{equation}
{\rm SFRD_{\rm III}} \propto M_{\rm III} \frac{\mathrm{d}}{\dt} N_{\rm h} = M_{\rm III} \frac{\mathrm{d}}{\dt} \int_{M_{\rm min}}^{M_{\rm a}} \mathrm{d}M \frac{\mathrm{d}n}{\mathrm{d}M}.
\end{equation}

To calculate the LW background which sets the minimum halo mass $M_{\rm min}$ for Pop-III star formation, we follow \citet{visbal15} and assume that at a given redshift $z$, the LW background receives contributions from all photons produced out to $z_e = 1.015z$, leading to
\begin{equation}
J_{\rm LW} = \frac{cZ^3}{4\pi} \int_{z_e}^z \mathrm{d}z' \frac{\dt}{\mathrm{d}z'} \left( \frac{{\rm SFRD_{III}}}{m_{\rm p}} N_{\rm LW,III} + \frac{{\rm SFRD_{II}}}{m_{\rm p}} N_{\rm LW,II} \right) \frac{h\nu_{\rm LW}}{\Delta\nu_{\rm LW}},
\end{equation}
where $Z\equiv1+z$ and $N_{\rm LW,III}=100,000,\ N_{\rm LW,II}=9690$ are the numbers of LW photons produced per stellar baryon in Pop-III and Pop-II stars respectively, $h\nu_{\rm LW} = 1.9\times10^{-11}$~erg and $\Delta\nu_{\rm LW} = 5.8\times10^{11}$~Hz.  The \citet{machacek01} model then gives
\begin{equation}
M_{\rm min} = 2.5\times10^5 \left( \frac{1+z}{26} \right)^{-1.5} \left( 1 + 6.96 \left( 4\pi J_{\rm LW}(z) \right)^{0.47} \right),
\end{equation}
where $J_{\rm LW}$ is in units of $10^{-21}$~erg/s/cm$^2$/Hz/Sr.

In addition to LW feedback, we assume that the Pop-III SFR is suppressed by photoheating feedback by a factor of $(1-\xion)$ \citep{irsic20}.  However, the effect of photoheating feedback on the resulting Pop-III SFR is very minor owing to the small ionized fraction (1-10\%) that Pop-III stars can produce at $z>15$.  %This assumes that the mass-averaged and volume-averaged ionized fractions roughly equal and that the ionization bubbles created by Pop-III star-forming halos are mostly isolated.

Table~\ref{tab:params} lists all the parameters we have used in the modeling and the values or ranges adopted.

\begin{table*}
\centering
\caption{Parameters used in our calculation of the reionization history of the universe and the corresponding values or ranges.}
\label{tab:params}
\begin{tabularx}{\textwidth}{l|X|X}
name & meaning & value/range \\
\hline
$f_{\rm III}$ & Pop-III star formation efficiency in the fiducial model assuming ${\rm SFRD_{III}} \propto f_{\rm III}\mathrm{d}\fcoll/\dt$ & $10^{-4}-10^{-2}$ \\
$M_{\rm III}$ & Pop-III stellar mass per halo in models assuming ${\rm SFRD_{III}} \propto M_{\rm III}\mathrm{d}N_{\rm h}/\dt$ & $100-500\ M_\odot$ \\
$M_{\rm min}(z, J_{\rm LW})$ & minimum halo mass of Pop-III star formation as a function of $z$ and $J_{\rm LW}$ & \citet{machacek01} (fiducial), \citet{trenti09, schauer20, kulkarni20} \\
$f_{\rm II}$ & Pop-II star formation efficiency & taken from \citet{tacchella18} (fiducial) and \citet{furlanetto17}, calibrated using UV luminosity functions and star formation rate density \\
$N_{\rm LW, III}$ & number of LW photons produced per stellar baryon in Pop-III stars & $100,000$ \\
$N_{\rm LW, II}$ & number of LW photons produced per stellar baryon in Pop-II stars & $9690$ \\
$N_{\rm ion, III}$ & number of ionizing photons produced per stellar baryon in Pop-III stars & $60,000$ \\
$N_{\rm ion, II}$ & number of ionizing photons produced per stellar baryon in Pop-II stars & $4000$ \\
$f_{\rm esc, III}$ & escape fraction of ionizing photons of Pop-III star-forming halos & 1 \\
$f_{\rm esc, II}$ & escape fraction of ionizing photons of Pop-II star-forming halos & 0-10 for models using the \citet{tacchella18} (fiducial) Pop-II model, 0-1 for models using the \citet{furlanetto17} Pop-II model \\
$C$ & IGM clumping factor & 1, 3 (fiducial), 5, 10
\end{tabularx}
\end{table*}

\section{LW feedback}
\label{sec:appendix_LW}

\begin{figure}
\centering
\includegraphics[width=\columnwidth]{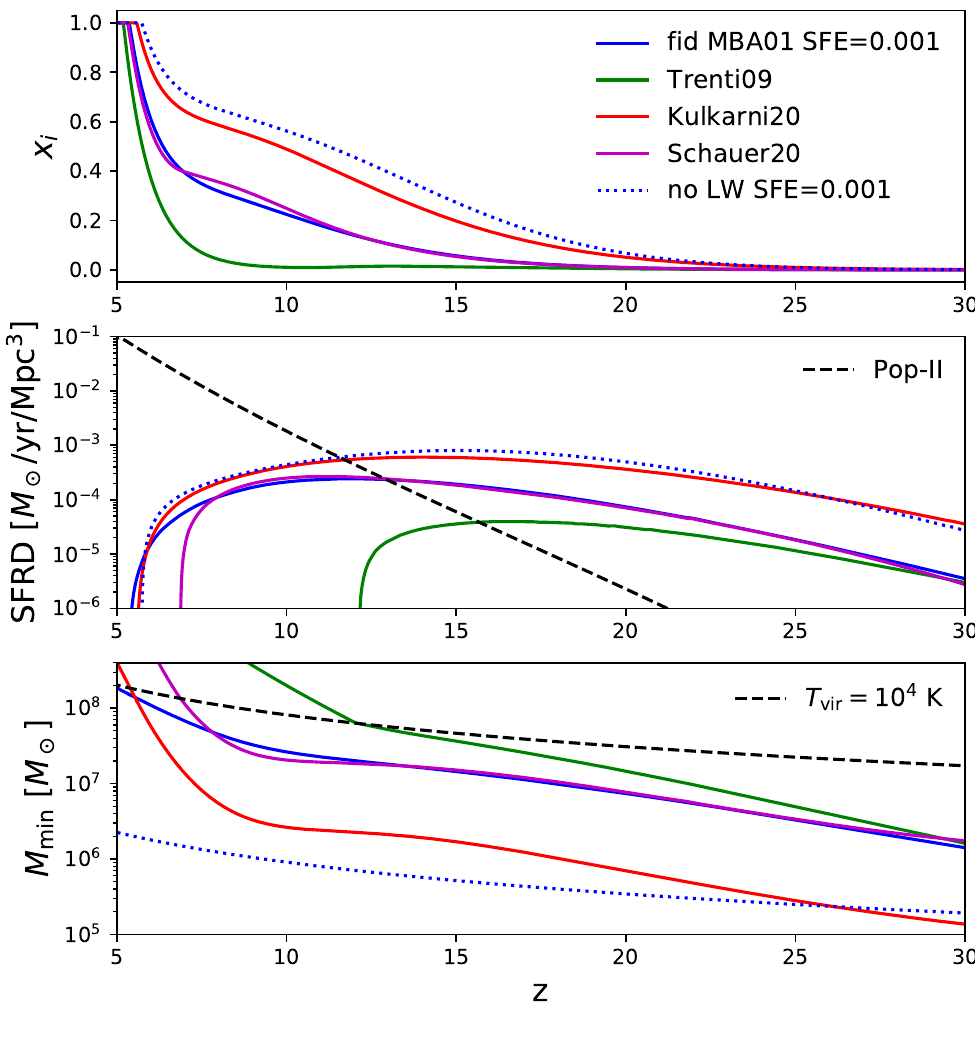}
\caption{Effects of different forms of LW feedback on the resulting ionized fraction evolution (top panels), Pop-III SFR density (middle), and minimum halo mass for Pop-III star formation (bottom).  Blue, green, red, and magenta represent the LW feedback models of \citet{machacek01}, \citet{trenti09}, \citet{kulkarni20}, and \citet{schauer20} respectively.  The calculations use the ``\dfcolldt'' Pop-III model with SFE $0.001$ and the Pop-II model of \citet{tacchella18} with $f_{\rm esc,II}=0.1$, and the Pop-II SFR density is shown as the black dashed line in the middle panels.  The black dashed line in the bottom panels illustrate halo mass corresponding to a virial temperature of $10^4$~K, above which we assume that atomic cooling dominates.  The blue dotted lines in both panels represent corresponding modeling results without LW feedback, assuming Pop-III stars form in halos with $T_{\rm vir} \ge 500$~K.}
\label{fig:model_Sandro_LW}
\end{figure}

We examine how different forms of LW feedback affects our calculation of the Pop-III SFR and reionization history.  Figure~\ref{fig:model_Sandro_LW} shows the modeling results of the ionized fraction (top), Pop-III SFR density (middle), and $M_{\rm min}$ (bottom), obtained using the functional form of $M_{\rm min}$ given by \citet{machacek01} (blue), \citet{kulkarni20} (red), \citet{schauer20} (magenta), and \citet{trenti09} (green).  The calculations use the ``\dfcolldt'' model with $f_{\rm III}=0.001$, and the Pop-II model of \citet{tacchella18} with $f_{\rm esc,II}=0.1$.  The Pop-II SFR density is shown as the black dashed line in the middle panel.  The black dashed line in the bottom panel illustrate halo mass corresponding to $T_{\rm vir}=10^4$~K, above which we assume that atomic cooling becomes efficient and star formation transforms to the Pop-II channel.  The blue dotted lines represent our calculations without LW feedback, so that Pop-III stars form in halos with $T_{\rm vir}\ge500$~K.

The \citet{kulkarni20} LW feedback model is the weakest among all the models, and acts similarly as our no LW case assuming Pop-III stars form in all halos with $T_{\rm vir}\ge500$~K.  The resulting $M_{\rm min}$ is only a factor of 3 higher than $T_{\rm vir}=500$~K.  We note that in \citet{kulkarni20} they found a $(1+z)^{-1.58}$ scaling of $M_{\rm min}$ without LW feedback, consistent with the redshift scaling of halo mass with a fixed virial temperature, but the amplitude is a factor of $\sim1.5$ lower than our assumed halo mass corresponding to $T_{\rm vir}=500$~K.  The \citet{kulkarni20} model thus results in high total $\tau$ of $0.091$.  The \citet{schauer20} $M_{\rm min}$ does not have redshift dependence and is roughly a factor of 2 smaller than the \citet{machacek01} $M_{\rm min}$ at $z=15$.  This model produces similar modeling results as the \citet{machacek01} model.  The \citet{trenti09} model is the strongest LW feedback model among all, leading to flatter Pop-III star formation histories and a factor of $\sim6$ lower Pop-III SFR at $z=15$ compared to the \citet{machacek01} model.  The \citet{trenti09} model also shows a saturation of the Pop-III star formation owing to the high amplitude and steep $(1+z)^{-3.6}$ scaling of $M_{\rm min}$, as $M_{\rm min}$ approaches the atomic cooling threshold at around $z=15$.  Such strong LW feedback also leads to negligible contribution of Pop-III stars to reionization, with $\tau(z>15)<0.001$.  In Section~\ref{sec:exotic_models} we thus lowered the amplitude of the \citet{trenti09} model to allow for higher $\tau(z>15)$.

%%%%%%%%%%%%%%%%%%%%%%%%%%%%%%%%%%%%%%%%%%%%%%%%%%

% Don't change these lines
\bsp	% typesetting comment
\label{lastpage}
\end{document}